\documentclass[draft]{agujournal2019}

\usepackage{graphicx}
\usepackage{epstopdf}
\usepackage{natbib}

\usepackage{url} 
\usepackage[inline]{trackchanges} 
\usepackage{soul}

\usepackage{fancyhdr}   
 \fancypagestyle{specialfooter}{%
 \fancyhf{}
 
 \fancyfoot[C]{Distribution Statement A.  Approved for public release.  Distribution unlimited.}  }

\journalname{Space Weather}

\begin{document}

\title{Forecasting the Remaining Duration of an Ongoing Solar Flare}

\authors{Jeffrey W. Reep\affil{1}}
\authors{Will T. Barnes\affil{2}}

\affiliation{1}{Space Science Division, US Naval Research Laboratory}
\affiliation{2}{National Research Council Research Associate Residing at the Naval Research Laboratory}

\correspondingauthor{Jeffrey Reep}{jeffrey.reep@nrl.navy.mil}



\begin{keypoints}
\item A random forest regression model has been trained on GOES-15 XRS data to predict the amount of time for an ongoing flare to return to background flux levels
\item The random forest prediction out-performs simple linear regression, and predicts the approximate end time with an average error of less than 2 minutes
\item This forecast can be run in real-time using only GOES/XRS data in order to forecast how long radio communications will be impacted
\end{keypoints}
\thispagestyle{specialfooter}   

\begin{abstract}
The solar X-ray irradiance is significantly heightened during the course of a solar flare, which can cause radio blackouts due to ionization of the atoms in the ionosphere.  As the duration of a solar flare is not related to the size of that flare, it is not directly clear how long those blackouts can persist.  Using a random forest regression model trained on data taken from X-ray light curves, we have developed a direct forecasting method that predicts how long the event will remain above background levels.  We test this on a large collection of flares observed with GOES-15, and show that it generally outperforms simple linear regression, giving a median error of less than 2 minutes for the approximate end time of a flare.  This random forest model is computationally light enough to be performed in real time, allowing for the prediction to be made during the course of a flare.
\end{abstract}

\section*{Plain Language Summary}
The X-ray emissions from a solar flare can impact the Earth's ionosphere, ionizing the atoms and thereby increasing the total electron content.  This is turn reduces the range of radio communications, effectively causing a blackout.  Unfortunately, it is not clear how long any given flare might last, and therefore how long communications could be adversely impacted.  The duration of a solar flare is not related to the size of that flare (as measured by the peak brightness of X-ray emissions), so it is not straightforward to predict how long it might last from simple observations.  In this work, we develop a method that allows us to forecast that duration in real time using a machine learning algorithm.  This allows us to predict how long radio communications will be impacted.  

\section{Introduction}

Solar flare emissions have been typically measured and classified by their soft X-ray emissions as measured by the X-ray Sensors (XRS) on board the Geostationary Operational Environmental Satellites (GOES).  These satellites measure the X-ray emission in the 1--8\,\AA\ and 0.5--4\,\AA\ wavelength bands, where the peak level in the former is used as a flare classification with X-class corresponding to peak emission above $10^{-4}$ W m$^{-2}$, and M, C, B, and A classes in decreasing orders of magnitude.  The frequency distribution of flare sizes is a power law distribution with slope of approximately -2 \citep[e.g.][]{hudson1991}, so the largest flares are significantly rarer than smaller ones.  

Solar flare durations vary widely, and are not correlated with the GOES class of the flare.  In \citet{reep2019}, for example, it was shown that the GOES class is not correlated with the full-width-at-half-maximum (FWHM) in either of the 1--8 \AA\ or 0.5--4 \AA\ XRS wavelength bands, and furthermore that the distributions of FWHM are consistent with log-normal distributions, approximately ranging from tens of seconds to a few hours.  There were also no relations found between the GOES FWHM and the temperature, emission measure (EM), thermal energy, ribbon area, or active region area, though there may be a relation in large flares between the magnetic flux and FWHM \citep{reep2019} or the sunspot area and the duration \citep{harra2016}.  Because the FWHMs do not correlate with the flare class, one cannot estimate an approximate duration from the peak flux of that flare (or similarly from other simple parameters).  \citet{toriumi2017} did find a linear relation between the separation of the flare ribbon centroids and the FWHM or e-folding decay time in a set of large flares (above M5), but this cannot be easily measured in real time, nor done with flares occurring at the limb, and would need to be extended to smaller flares to be of general use.  

The D-region of the Earth's ionosphere is strongly impacted by solar flare emission (\textit{e.g.} \citealt{thomson2005}), particularly from Lyman-$\alpha$ \citep{milligan2020}, and X-rays during flares \citep[e.g.][]{levine2019}.  The solar emission ionizes molecular nitrogen (N$_{2}$) and oxygen (O$_{2}$), and the less abundant nitric oxide (NO), which in turn attenuates the propagation of high frequency (HF) radio waves.  Models have been developed to understand the impact of irradiance variability on ionospheric absorption and the attenuation of HF radio communications at various frequencies, such as the Data-Driven D Region model \citep[DDDR,][]{eccles2005}, NOAA's empirical D-Region Absorption Prediction \citep[D-RAP, based on][]{sauer2008} or the ray-tracing Modified Jones (MoJo) HF propagation code \citep{zawdie2015, zawdie2017}.  Prediction of the solar irradiance during the course of a flare is thus vital to predicting the impacts on radio communications.  Crucially, since the duration of a flare is independent of its magnitude, in long duration events, the total irradiant energy could be significant for hours beyond the peak of the event, thus causing a lasting effect on HF propagation.  It is therefore crucial to forecast the duration in order to properly understand the attenuation of HF radio waves.

It is not clear that the duration of a solar flare in X-rays can be predicted from simple scaling laws, however.  The duration may be related to some parameters that can be calculated \textit{post facto}, but this means that a real time forecast would be impossible.  However, given that there are thousands of flares observed by GOES of widely varying sizes and durations, and at various levels of solar activity, the statistical data should contain information to estimate likelihood of the duration.  This is the goal of this paper.  

\begin{figure*}
    \centering
    \includegraphics[width=\linewidth]{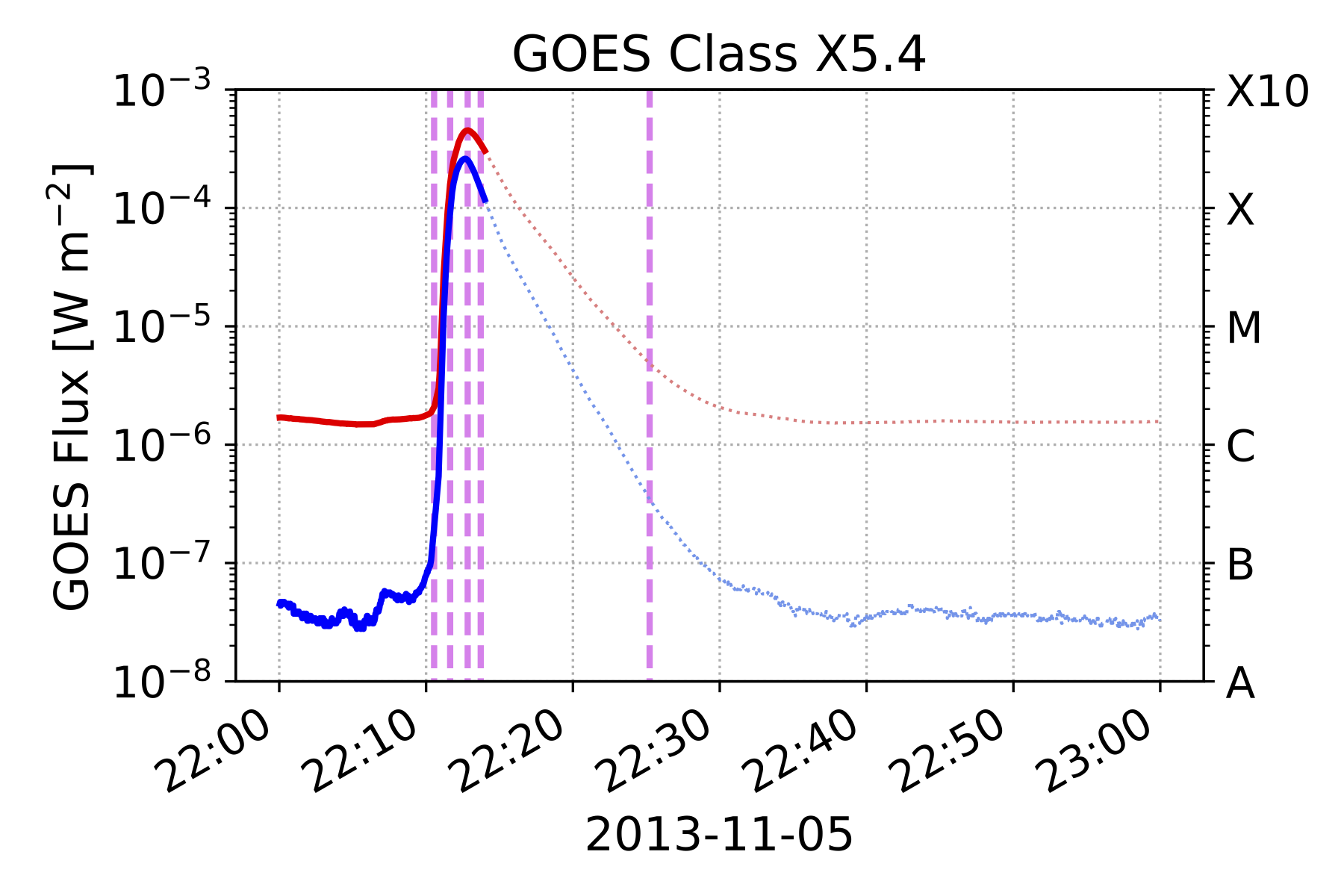}
    \includegraphics[width=\linewidth]{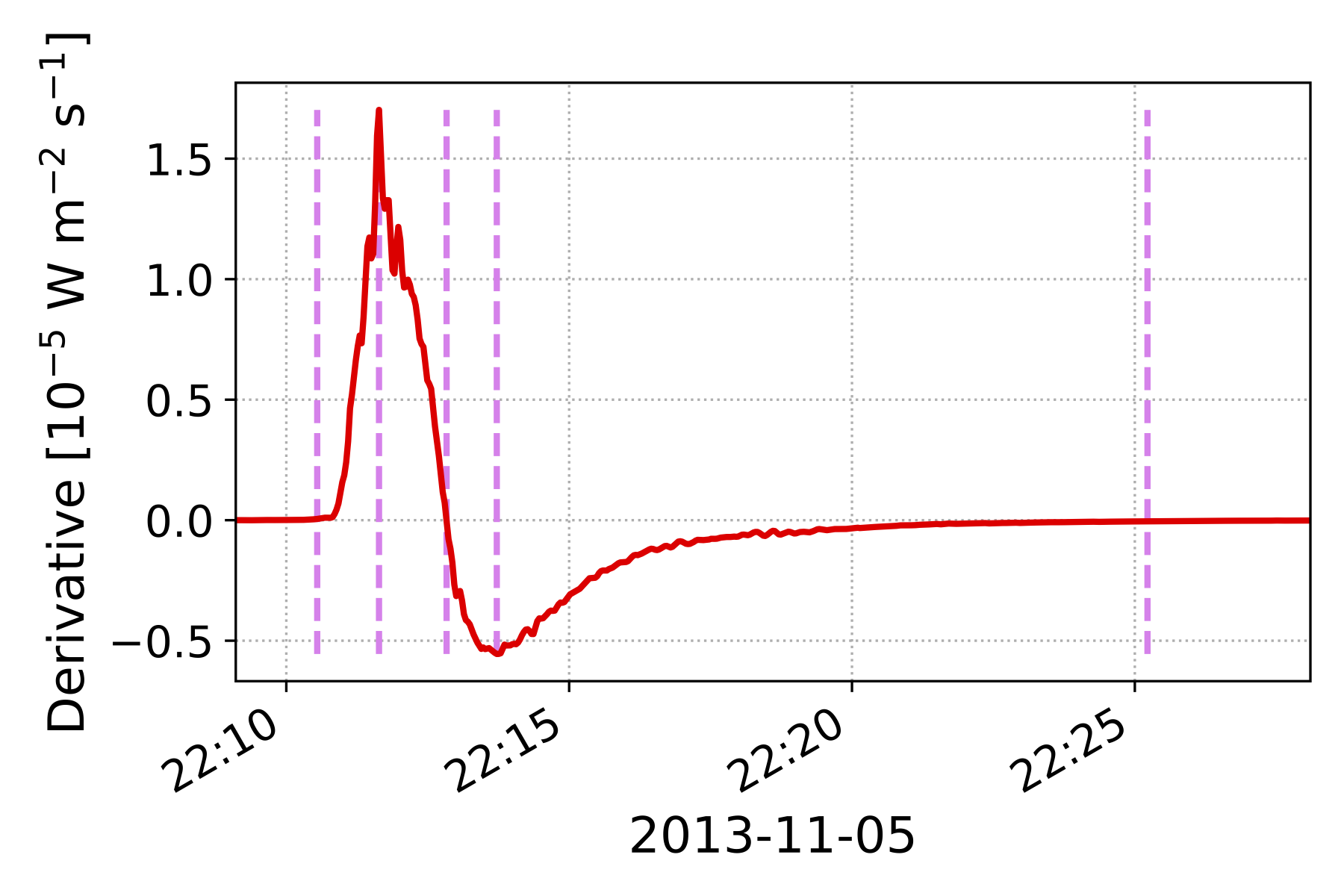}
    \caption{An example event where we wish to predict the approximate duration (see the text).  At top, the figure shows the GOES XRS-B 1--8 \AA\ light curve (red) and XRS-A 0.5--4 \AA\ light curve (blue) up to about UT 22:14 for an X5.4 flare on 5 November 2013 (solid lines), while the dotted lines show the true light curves.  The bottom plot shows the time derivative of XRS-B.  The five dashed vertical lines on each plot mark the values of each successive $t_{i}$ (see Section \ref{sec:data}).  }
    \label{fig:problem}
\end{figure*}

Therefore, we wish to develop a real-time forecasting tool that predicts how long a solar flare will take to cool from its current flux level to an approximate background level.  For example, suppose that an X-class flare like the one in Figure \ref{fig:problem} were occurring at the present moment.  The two GOES/XRS channels are shown (top) in red (XRS-B, 1--8 \AA) and blue (XRS-A, 0.5--4 \AA), as well as the derivative of the long wavelength channel (bottom).  Given the light curves up to the present and their current flux values (solid lines), could we estimate how long they would take to return to the background level?  What is the likely duration and what is the range of potential values?  In other words, how well can we forecast the true light curves (dotted lines)?  To do so, we use quantities measured from the light curves at five times $t_{i}$, denoted by vertical pink lines, defined by the derivatives, for the 1--8\,\AA\ channel. These are fully explained in Section \ref{sec:data}.  We attempt to predict the timing of the final vertical line, which we call $t_{4}$ in this paper.

In this work, we use a random forest regressor trained only with parameters measured from observed GOES/XRS data to estimate the approximate duration of the flare in XRS and calculate a likelihood interval of the duration.  We show that this model improves its prediction over the course of a flare, and that it outperforms a linear regression model.  

\section{Data Preparation}
\label{sec:data}

We first prepare data taken from XRS observations of flares with measurements made by the GOES-15 satellite, which had a 2.048 s time cadence.  We examine all flares in NOAA's GOES event catalogue from the launch of GOES-15 on 4 March 2010 through the deactivation of the satellite on 2 March 2020.  While we have chosen to only focus on GOES-15 data, the model could be trained with data from any GOES satellite.  Because the instrumental response varies slightly for each satellite, it is possible that predictions trained with one satellite's data may not work well for another satellite, which should be tested in the future.

We expand the time range of the events in the GOES catalogue to ten minutes earlier and sixty minutes later than the listed start and end times to better include pre-flare rise and late phase decay.  We do not background subtract the data, since we do not know \textit{a priori} whether the background activity impacts flare duration.  We remove events from the data set if any of the following apply to an event: (1) there are any data gaps in either XRS channel; (2) the signal-to-noise ratio in either XRS channel is less than 2; (3) the event is not sufficiently isolated in time from other events, which we explain below based on the timings; (4) basic parameters like the FWHM, fluence, and time derivative are well-defined at all times.

We define five times for each XRS channel (\textit{i.e.} ten total times) in terms of the derivative with respect to time of the light curves: $t_{0}$ the onset of the flare as the derivative begins to rise, $t_{1}$ when the derivative is maximized, $t_{2}$ when the flux peaks, $t_{3}$ when the derivative is minimized, and $t_{4}$ when the derivative approximately returns to 0.  More specifically, we define $t_{0}$ as the first time when the derivative exceeds $10^{-4}$ times the peak flux; for example, for an X1 flare, we define $t_{0}$ as the first time when the derivative exceeds $10^{-8}$ W m$^{-2}$ s$^{-1}$.  Note that we scale this threshold with flare class since, for example, the peak magnitude of the time derivative of a C-class flare might not even hit the threshold that defines the start time for an X1 flare, and the threshold for a small flare would be unnoticeable during active periods capable of producing large flares.  We similarly define $t_{4}$ as the first time after $t_{3}$ when the derivative is above $-10^{-4}$ times the peak flux; for an X1 flare, $t_{4}$ is the first time after $t_{3}$ when the derivative increases above $-10^{-8}$ W m$^{-2}$ s$^{-1}$.  We list the definitions in Table \ref{tab:ti}.  In the rest of this text, we refer to $t_{4} - t_{0}$ as the ``total duration'' of the event, and generally $t_{i} - t_{0}$ as a ``duration.''

For each flare, we measure 52 parameters for each event from the light curves.  We first calculate the derivatives in both channels using a 32-point Savitzky-Golay smoothing filter \citep{savitzky1964}.  The filter is particularly important for small flares where the signal-to-noise ratio is not as good as in larger flares.  

\begin{table}
\caption{Definitions of the flare timings that we use in this work.  The values are all implicitly defined in terms of the first time derivative of the light curves $F(t)$ of each XRS channel.  $t_{0}$ and $t_{4}$ are the first times for which the condition holds.  For all flares in our data set, we require $t_{i}$ $<$ $t_{i+1}$ to be certain that each event is isolated from other events in time.\label{tab:ti}}
\begin{tabular}{c | l | l}
    Time & Meaning & Definition \\ \hline
    $t_{0}$ & Approximate Start Time & $\frac{dF}{dt}|_{t_{0}} = (10^{-4}\ \mathrm{s}^{-1}) \times F(t_{2})$ \\ 
    $t_{1}$ & Maximum of $\frac{dF}{dt}$ & $\frac{dF}{dt}|_{t_{1}} = \max{\Big(\frac{dF}{dt}\Big)}$ \\ 
    $t_{2}$ & Peak of Flare & $\frac{dF}{dt}|_{t_{2}} = 0$ and $F(t_{2}) = \max{(F)}$\\
    $t_{3}$ & Minimum of $\frac{dF}{dt}$ & $\frac{dF}{dt}|_{t_{3}} = \min{\Big(\frac{dF}{dt}\Big)}$\\
    $t_{4}$ & Approximate End Time & $\frac{dF}{dt}|_{t_{4}} = -(10^{-4}\ \mathrm{s}^{-1}) \times F(t_{2})$
\end{tabular}
\end{table}

At each time $t_{i}$, we measure in both XRS channels the flux level $F_{i} = F(t_{i})$, the derivative of the flux with time, $\frac{dF}{dt}|_{t_{i}}$, and the flare fluence (definite integral) $A_{i} = \int_{t_{0}}^{t_{i}} F(t)\ dt$.  By definition, $A_{0} = 0$ and $\frac{dF}{dt}|_{t_{0}} \approx 0$, so these are neglected.  This gives 52 total parameters, 26 for each channel.  We train the random forest regressor twice: once using all the values through time $t_{2}$ in order to forecast $t_{3}$ and $t_{4}$ for each channel, and once using the values through time $t_{3}$ in order to forecast $t_{4}$.  Since we do not know \textit{a priori} what parameters act as strong predictors, we consider all features when training the random forest predictor.  In this work, we refer to the flux at time $t_{4}$, $F(t_{4})$, to be the ``background'' flux for simplicity in our terminology, but note that it is not a trivial problem to measure the true background level in XRS data (see discussion by \citealt{ryan2012}).

Because flares often occur within short time periods of one another, it can be difficult to measure some of the parameters with which we might wish to train a machine learning algorithm.  For example, if a second flare occurs before an earlier flare finishes and the X-ray emission has not yet returned to background levels, we cannot measure the true end time of the first flare. This means that we cannot include it in our training data set. Specifically, we enforce the condition that each $t_{i}$ occurs in sequential order, and discard events where this condition does not hold. This happens, for example, when there are flares of similar size occurring in close succession.  

\begin{figure}
    \includegraphics[width=\linewidth]{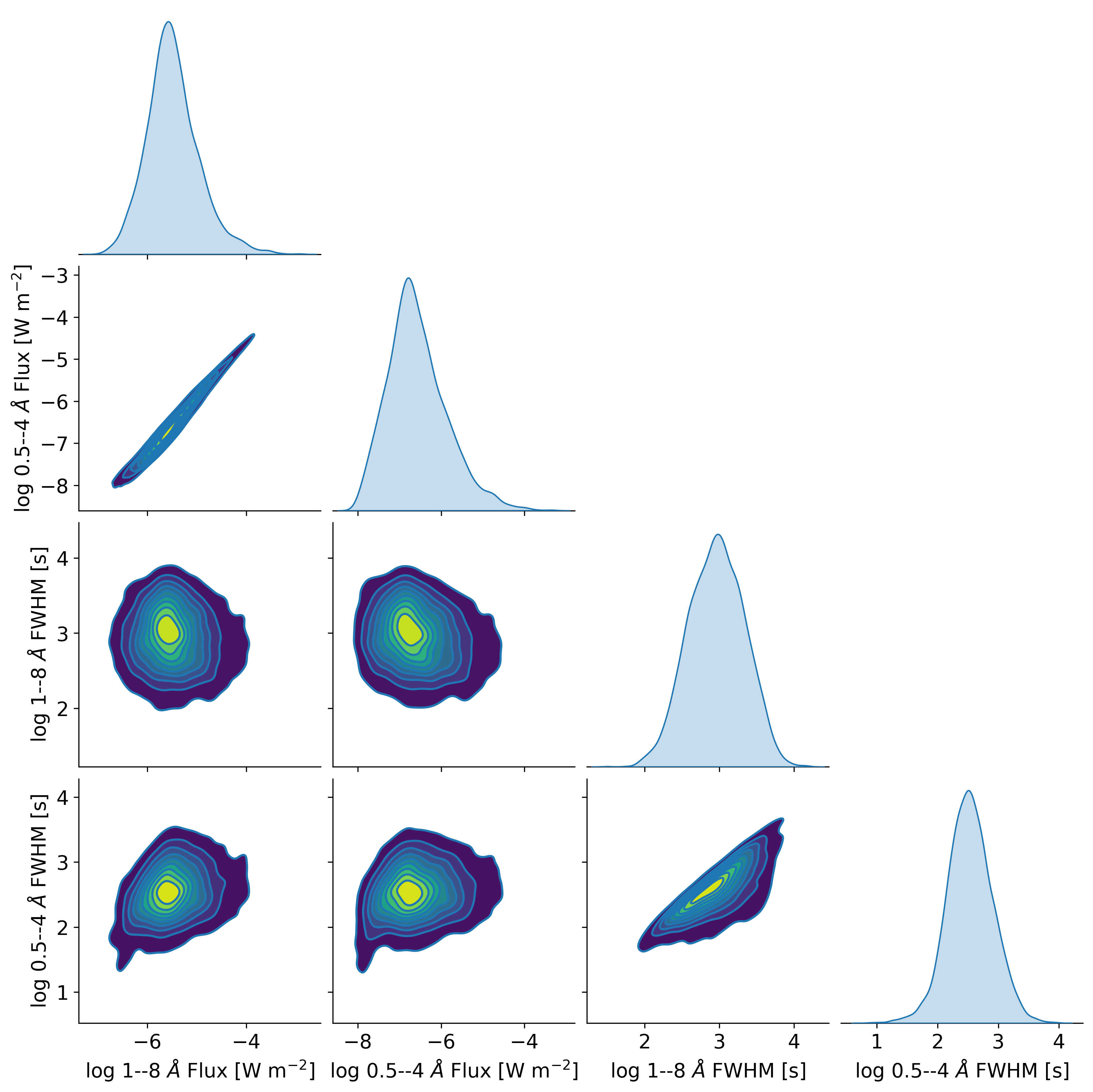}
    \caption{Plots of the distributions of the peak fluxes and FWHM in the two XRS channels.  The diagonal elements show plots of the relative frequencies, while off-diagonal elements show heat maps comparing each pair of variables.  The peak flux levels and FWHMs are uncorrelated, and each distribution is consistent with log-normal (see \citealt{reep2019}).}
    \label{fig:class_dur}
\end{figure}

After pruning the data set, we are left with 5730 events, ranging in GOES class from A9.8 to X13.  The FWHM of the light curves range between 30 and 16068\,s in the 1--8\,\AA\ channel and between 6 and 10307\,s in the 0.5--4\,\AA\ channel. The flares occurred between 4 September 2010 and 6 July 2019 across a wide range of solar activity levels.  In Figure \ref{fig:class_dur} we show the distributions of fluxes and FWHMs in both XRS channels, as well as heat maps showing the relations between each variable.  The fluxes and FWHMs are uncorrelated, while each individual distribution is consistent with log-normal, demonstrated with Kolmogorov-Smirnov tests in \citet{reep2019}.  

\begin{figure}
    \centering
    \includegraphics[width=\textwidth]{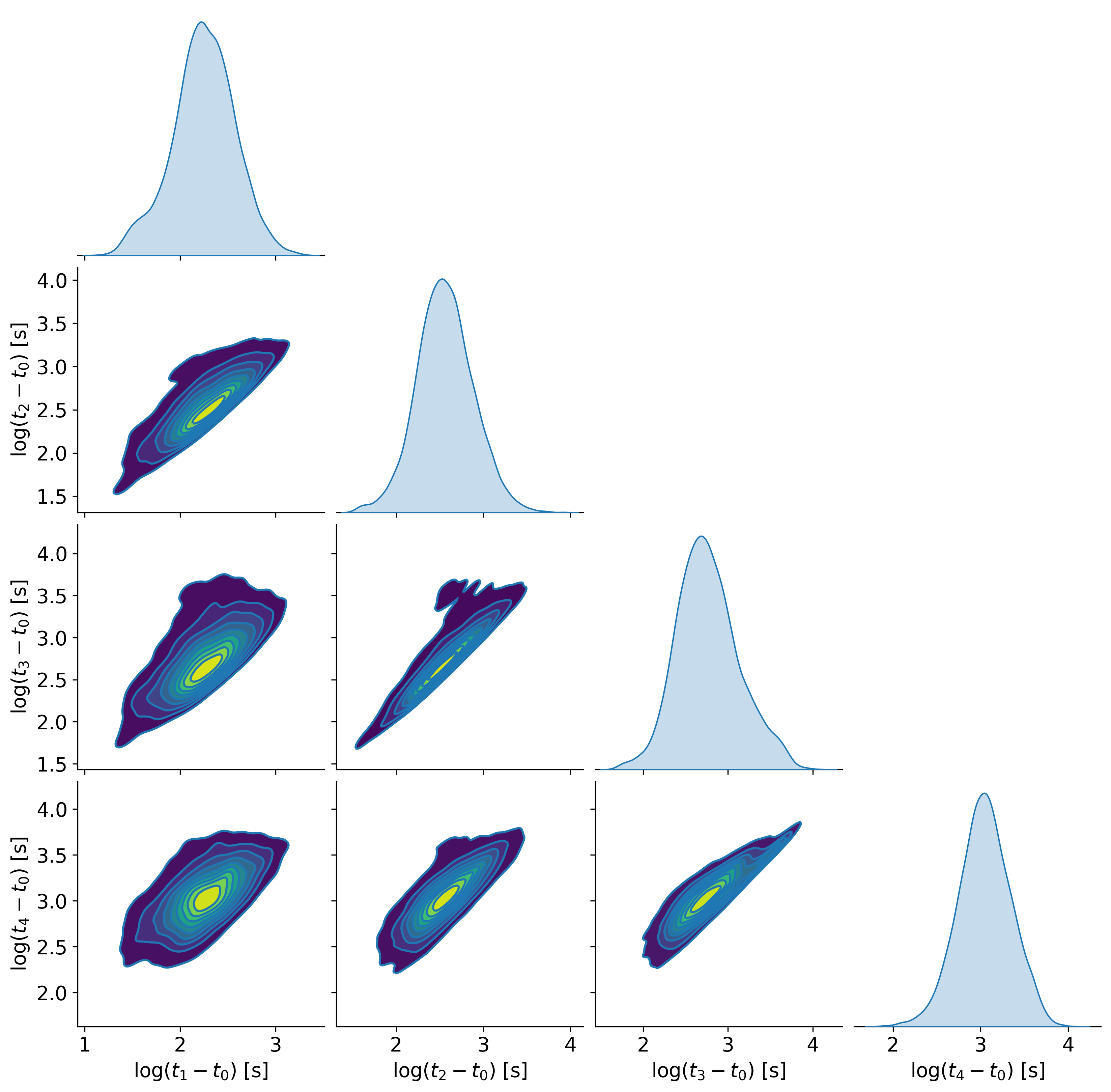}
    \caption{A pair plot showing the relationships between the timings $t_{i} - t_{0}$, for each $i > 0$, in the 1--8\,\AA\ channel.  While there is a general positive and monotonic correlation, there is also noticeable scatter.}
    \label{fig:ti_pairs}
\end{figure}

Similarly, Figure \ref{fig:ti_pairs} shows the relationships between the timings in the events in the 1--8\,\AA\ channel. We show each $t_{i} - t_{0}$, $i > 0$, which in all cases show a positive and monotonic correlation.  However, the scatter is noticeable, particularly when going from early timings to late ones.  It is clear that a linear regression could then give a prediction for $t_{3}$ and $t_{4}$, but would not capture the scatter.  Examining the last row, for example, it is clear that $t_{1}$ would not give a good prediction for $t_{4}$, but $t_{2}$ would give a better prediction, and $t_{3}$ better still.  We show in the next section that a random forest regressor outperforms a simple linear case.

\section{Duration Prediction}

We use the random forest regressor implemented in the \texttt{scikit-learn} package \citep{scikit-learn} to calculate this prediction.  Random forest methods \citep{breiman2001} create an ensemble of decision tree predictors, which when averaged give a robust prediction.  The chief advantage of a random forest predictor is that it does not assume any functional form for the data being fit (whether through classification or regression), so that the model can map complex relationships between the inputs and outputs.  \citet{breiman2001} notes that the other main advantages are that random forests are fast, robust against noise, and can be used to quantify errors and correlations easily.  

Decision trees \citep{breiman1984} work by recursively splitting a data set, based on values of training features, into nodes.  The trees split a certain number of times (the depth), which can be fixed or randomized.  While an individual decision tree is not robust in predictions, by combining multiple trees with differing nodes and depths, the prediction is improved.  The random forest predictor randomizes the elements of the set that are used in a decision tree, the features used in the splitting at each node, the values by which they are split, and the depth of the trees. When combined, these many varying decision tree predictors significantly reduce the variance of the model.  

\subsection{Predictions at time $t_{2}$}
We separate the full data set randomly into a training group (67\% of events) and test group (33\% of events), using the features of the light curves defined above to train the data.  We train the random forest regressor with the former group.  We then test the efficacy of the model by predicting $t_{3}$, $t_{4}$, $A_{3}$, and $A_{4}$ for the remaining 33\% of the events and comparing them to the actual, observed values.

\begin{figure*}
\centering
\includegraphics[width=0.48\textwidth]{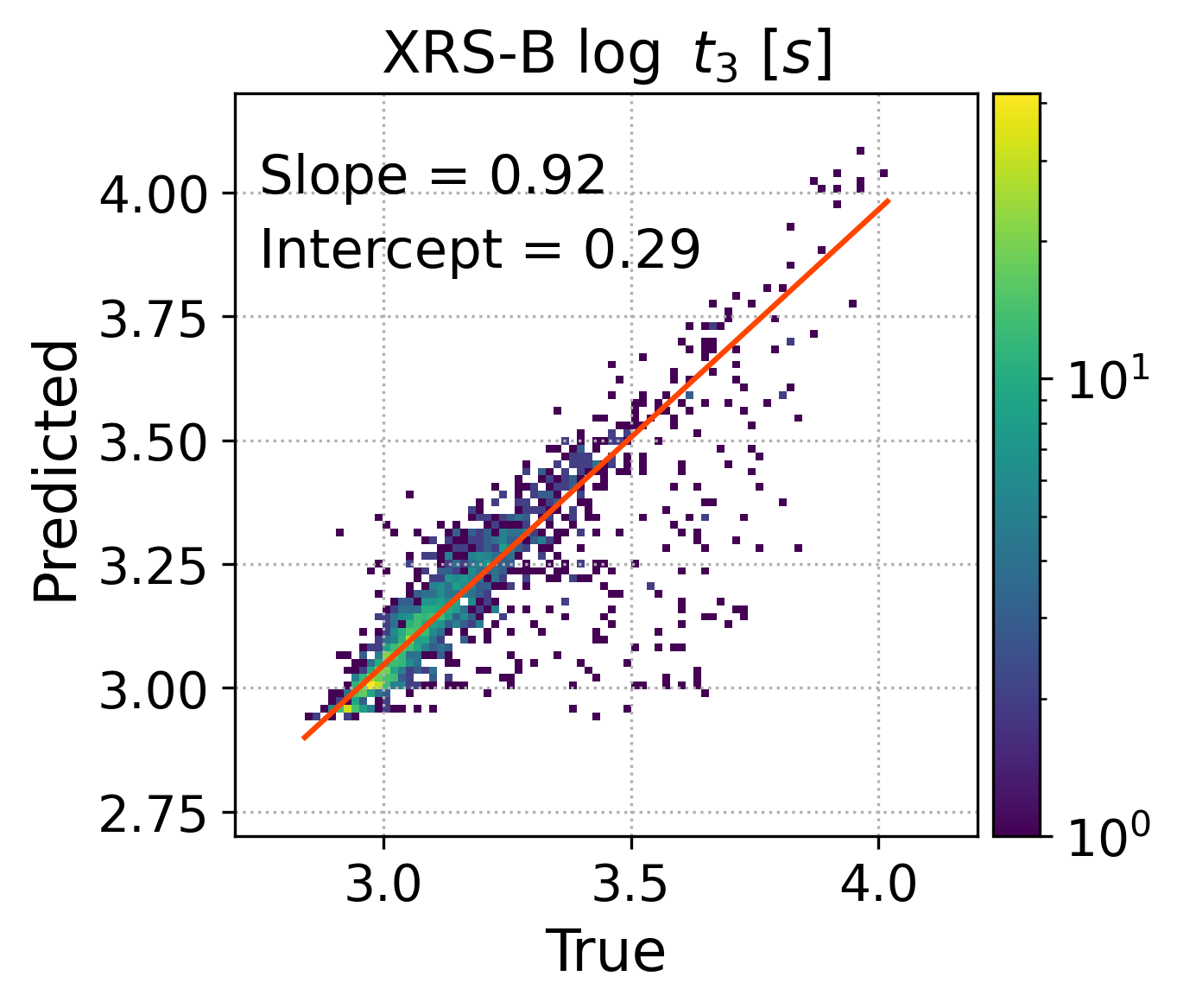}
\includegraphics[width=0.48\textwidth]{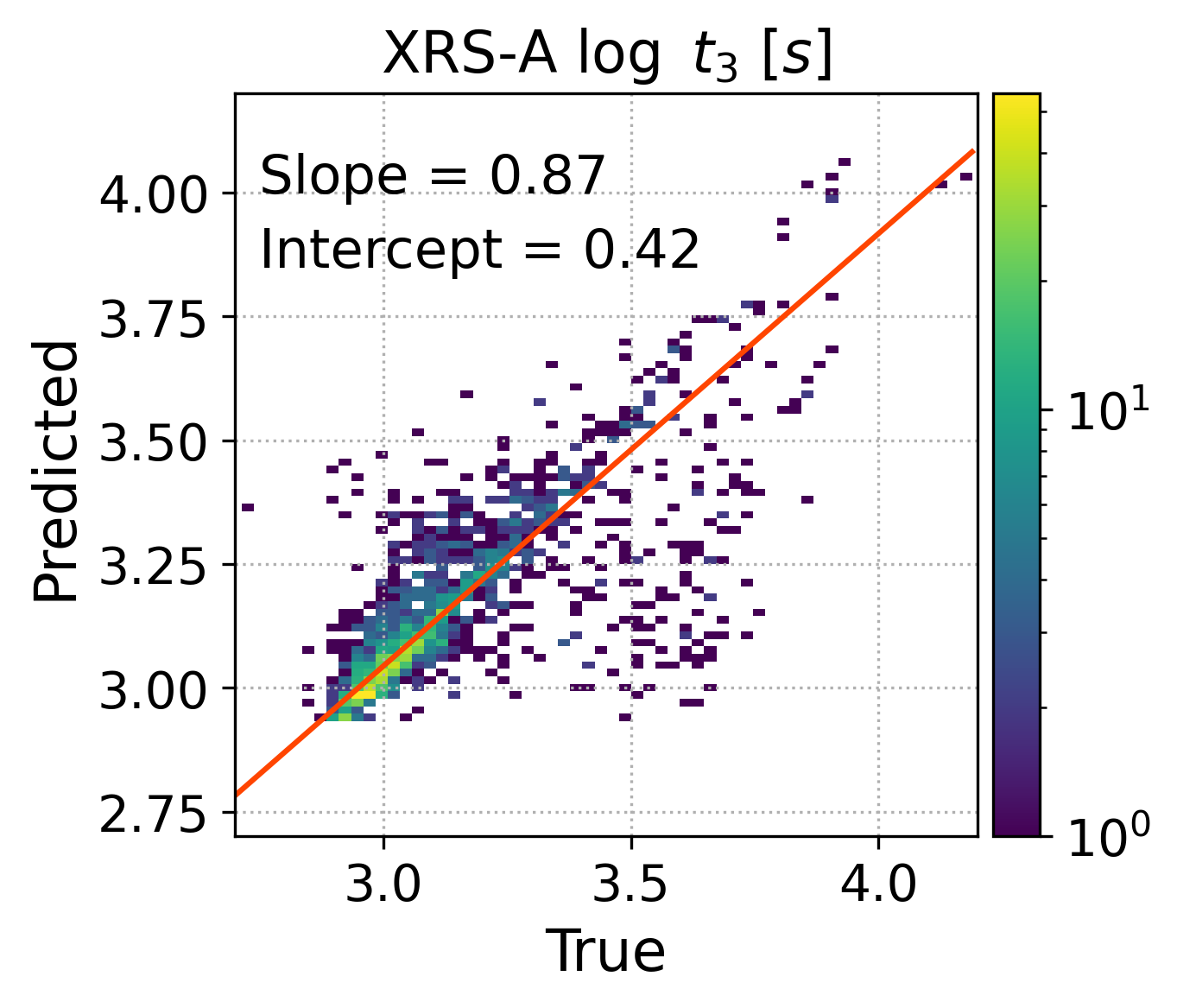}
\includegraphics[width=0.48\textwidth]{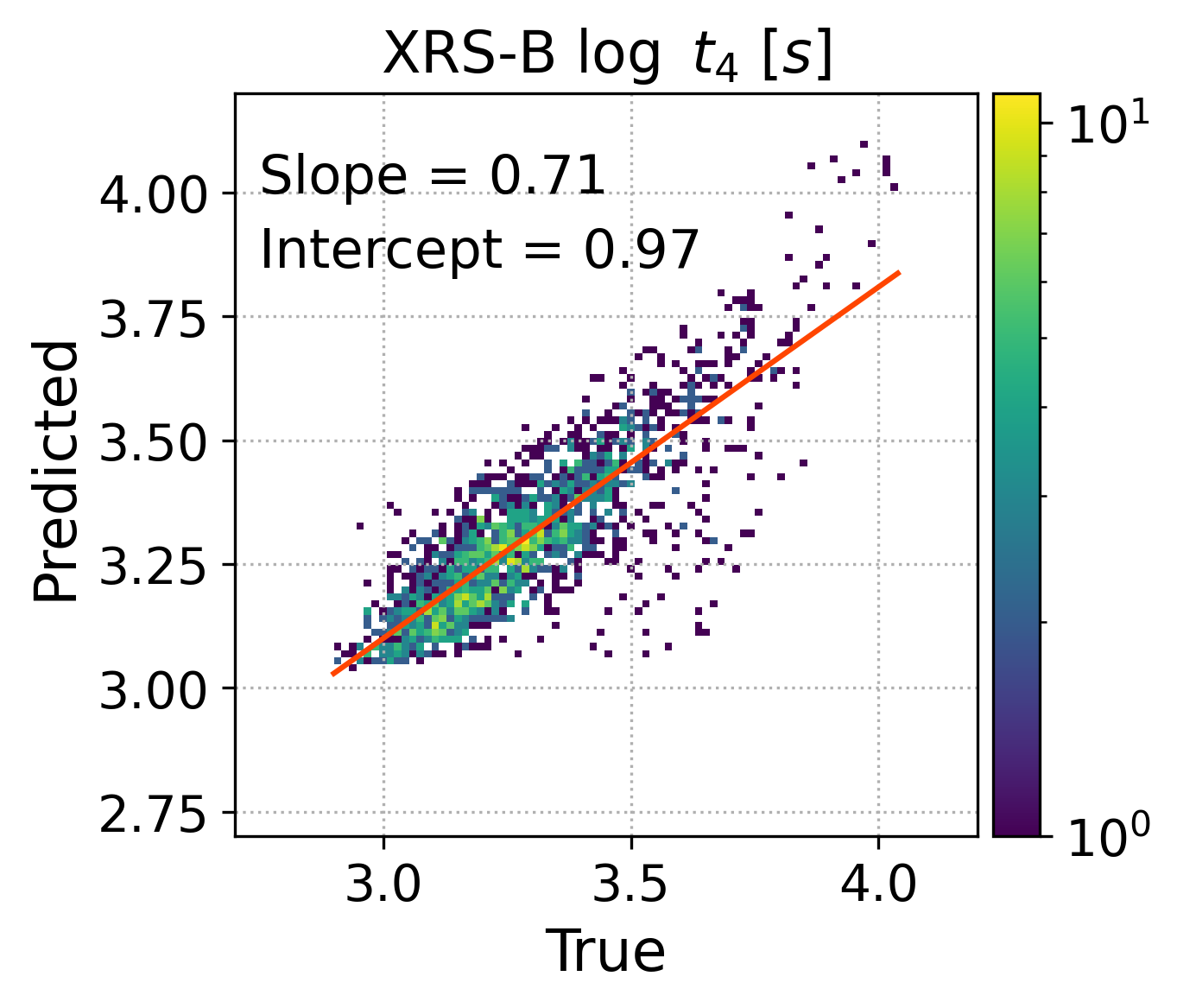}
\includegraphics[width=0.48\textwidth]{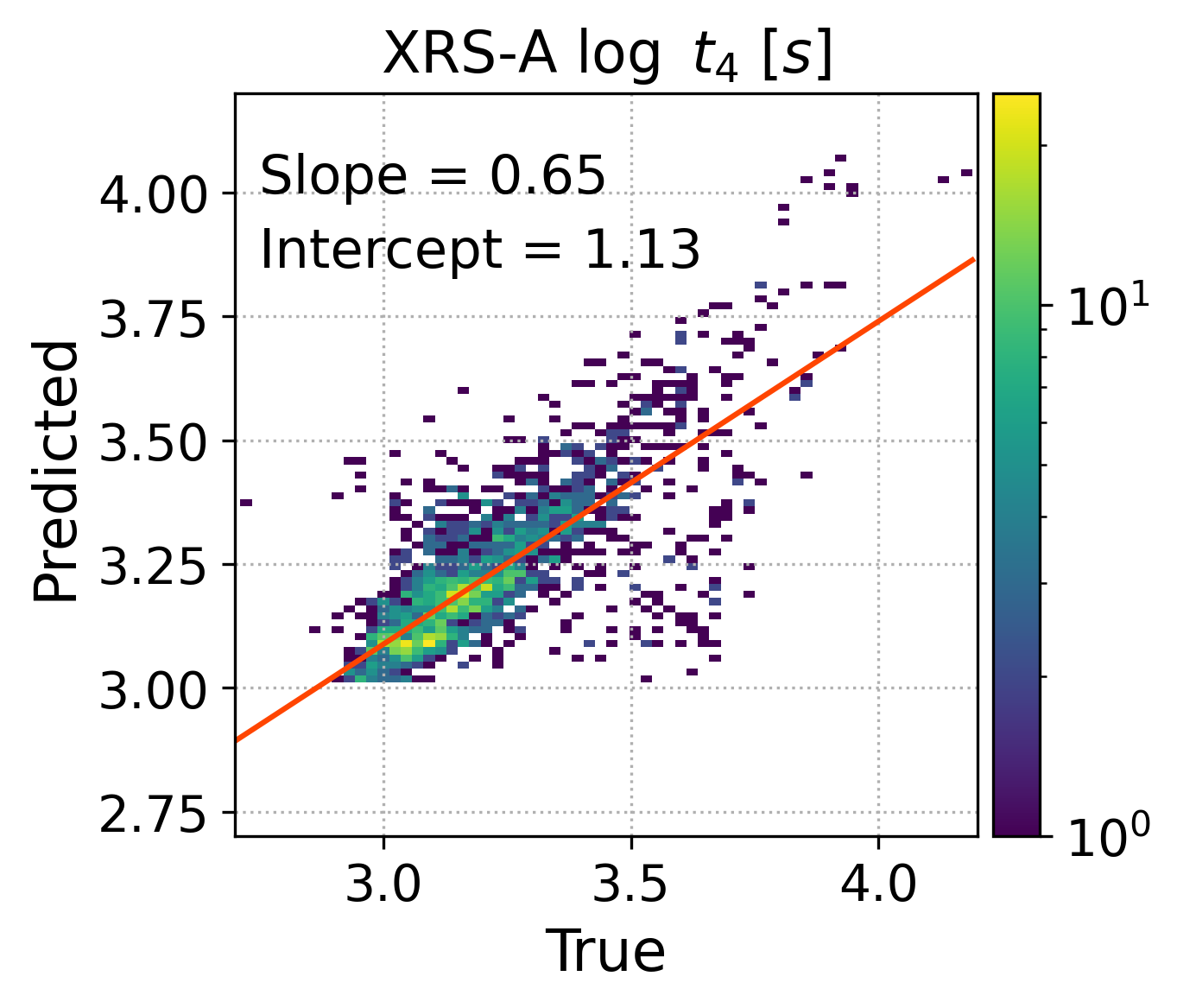}
\includegraphics[width=0.48\textwidth]{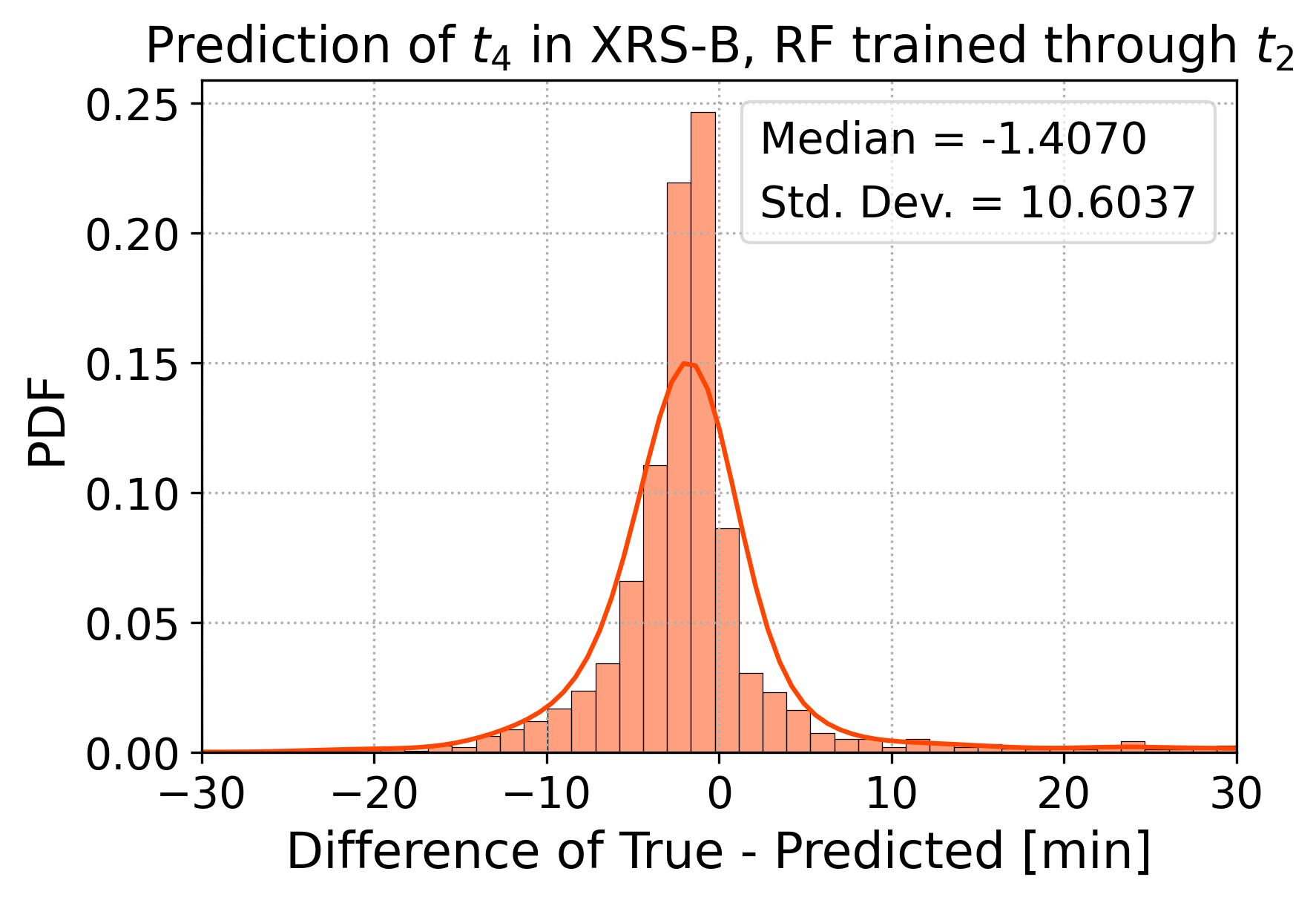}
\includegraphics[width=0.48\textwidth]{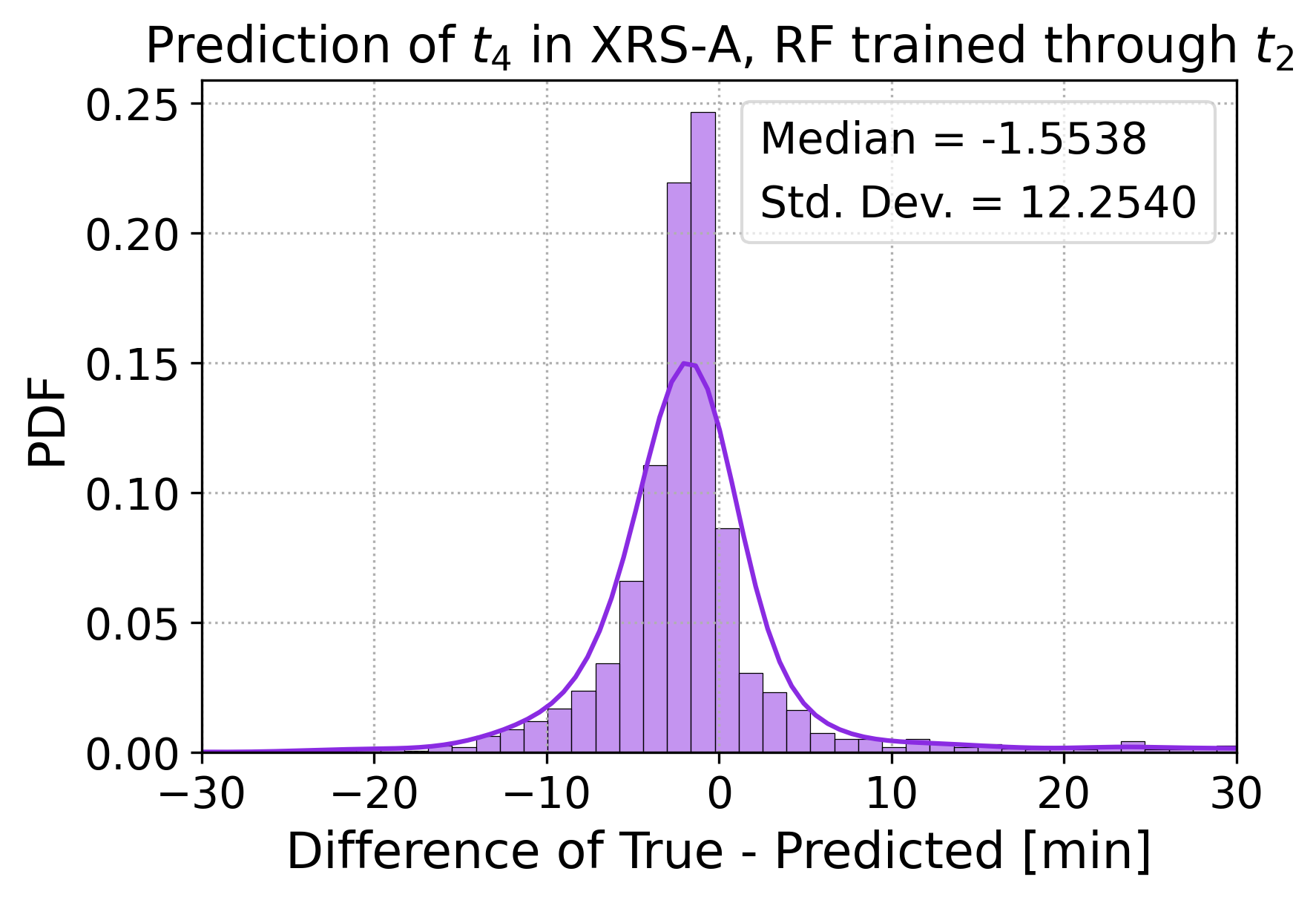}
\caption{One random forest regressor prediction of the remaining $t_{3}$ and $t_{4}$, using a forest trained with the parameters through time $t_{2}$.  The plots show 2D histograms comparing the predicted and true values of $t_{3}$ (top row) and $t_{4}$ (bottom row) for the XRS 1--8\,\AA\ (left column) and 0.5--4\,\AA\ (right column) channels for the 33\% of events placed in the test group.  The 
were performed using only the features through time $t_{2}$.  The orange lines are Theil-Sen linear regression fits to the data, with the slope and intercepts indicated.  The slopes are less than 1, indicating that the predicted values are underestimated in general.  At bottom, we show the probability density functions (PDFs) of the error in the predictions of $t_{4}$, with median errors of about -1.5 and -1.9 minutes.}
\label{fig:predt2_t3t4}
\end{figure*}

These predictions are presented in Figure \ref{fig:predt2_t3t4} for the test group.  Using the parameters through time $t_{2}$ to train the random forest, we show the estimated values of $t_{3}$ (top row) and $t_{4}$ (bottom row) for XRS-B (left column) and XRS-A (right column) as compared to their true values.  We perform a linear regression with a non-parametric Theil-Sen estimator \citep{sen1968,theil1992} to estimate the slope and intercept for these scalings.  Theil-Sen estimators are robust and insensitive to outliers in the data, unlike ordinary least squares linear regression.  We would expect a slope of 1 for a perfect prediction, such that the predicted time grows at the same rate.  The predictions for $t_{3}$ scale approximately with a slope of 0.9, while the predictions for $t_{4}$ scale more slowly, suggesting that the forecast will generally underestimate the $t_{4}$ for longer events.  The probability density functions (PDFs) at the bottom of the figure show the distribution of true minus predicted values of $t_{4}$, showing that on average $t_4$ in XRS-A and XRS-B is underestimated by approximately -1.5 and -1.9 minutes, respectively.

\begin{figure*}
\centering
\includegraphics[width=0.48\textwidth]{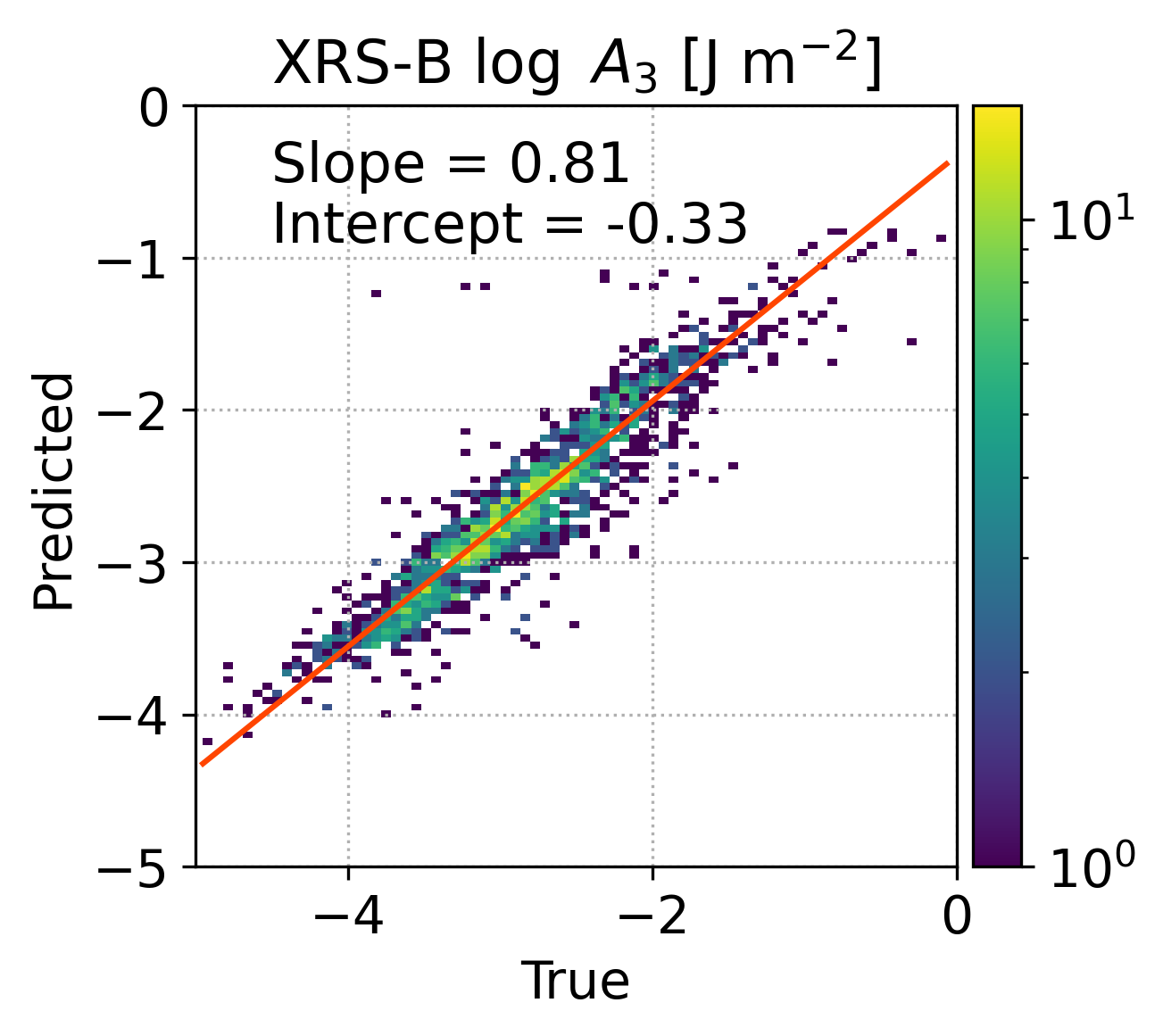}
\includegraphics[width=0.48\textwidth]{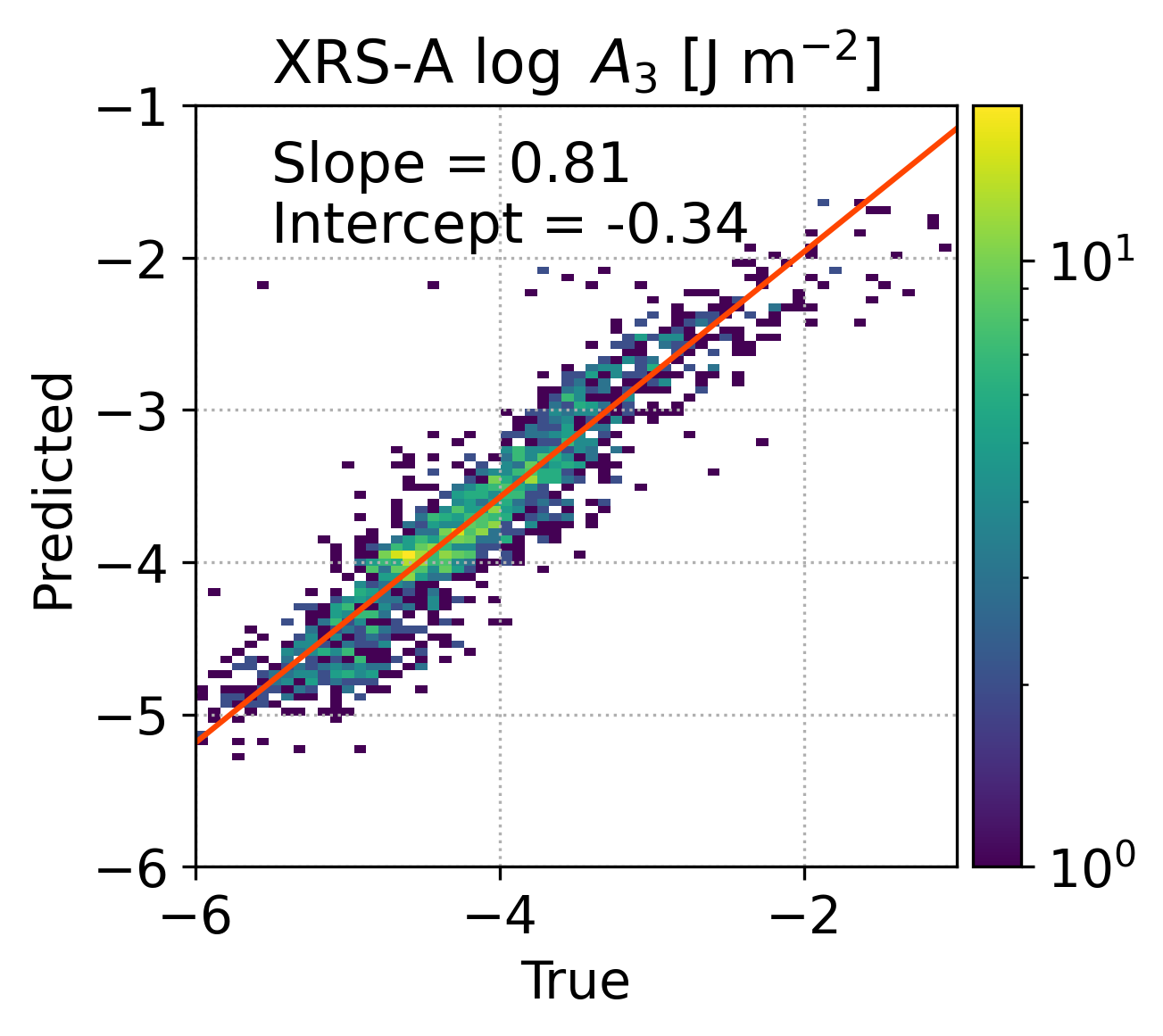}
\includegraphics[width=0.48\textwidth]{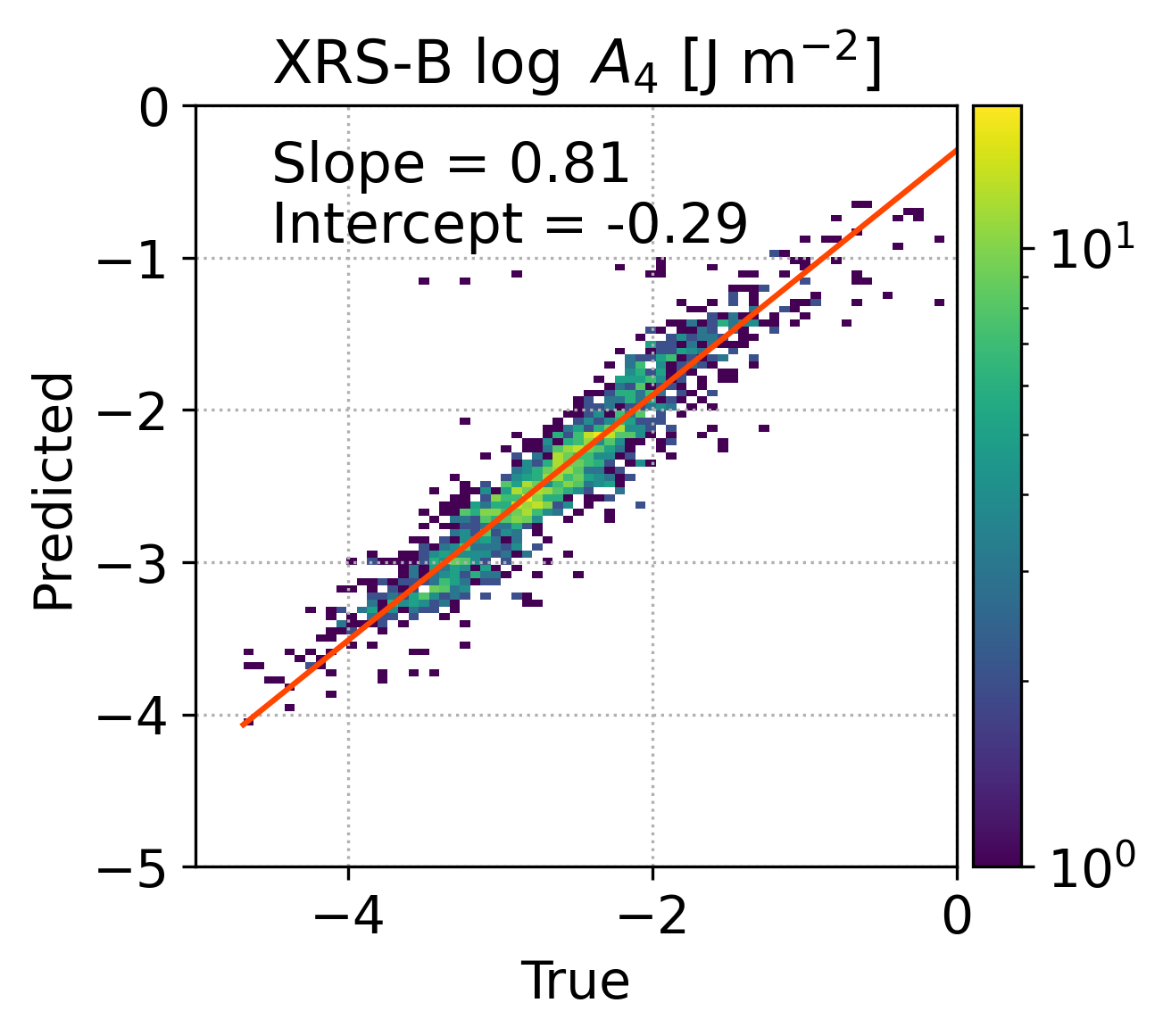}
\includegraphics[width=0.48\textwidth]{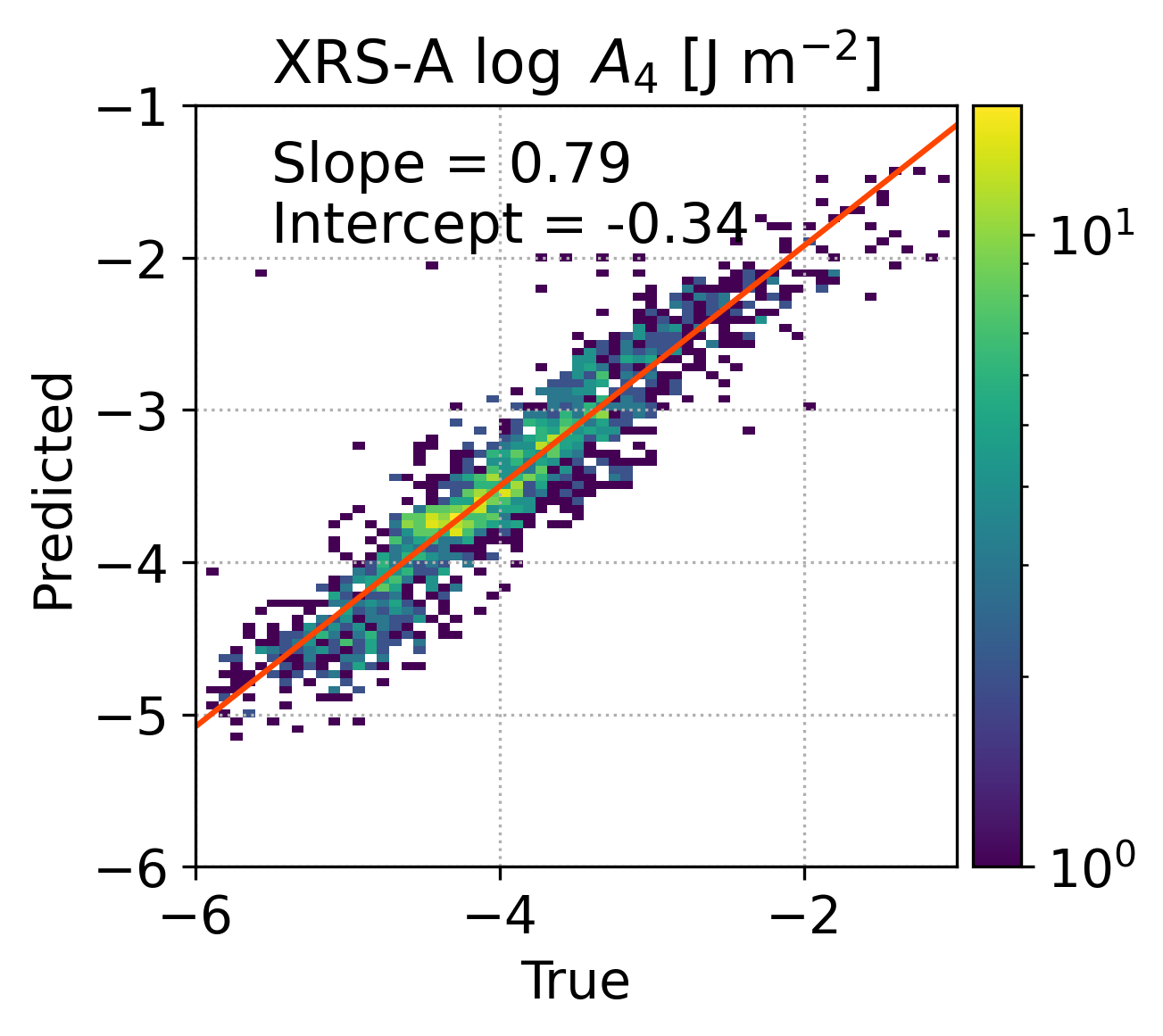}
\caption{A random forest prediction of the fluences $A_3$ and $A_4$ in both XRS channels, with the random forest trained through time $t_{2}$.  As with the timings, the predicted fluences are generally underestimated (slope less than 1).}
\label{fig:predt2_A}
\end{figure*}

We additionally show in Figure \ref{fig:predt2_A} the predictions of the fluences $A_{3} = \int_{t_{0}}^{t_{3}} F(t) dt$ and $A_{4} = \int_{t_{0}}^{t_{4}} F(t) dt$ in both XRS channels.  In this case, the slopes are all around 0.8, suggesting that the fluences at each time are predicted with similar accuracy, but once again are less than 1, so the predicted fluences are somewhat underestimated on average.

\begin{figure*}
\includegraphics[width=0.5\textwidth]{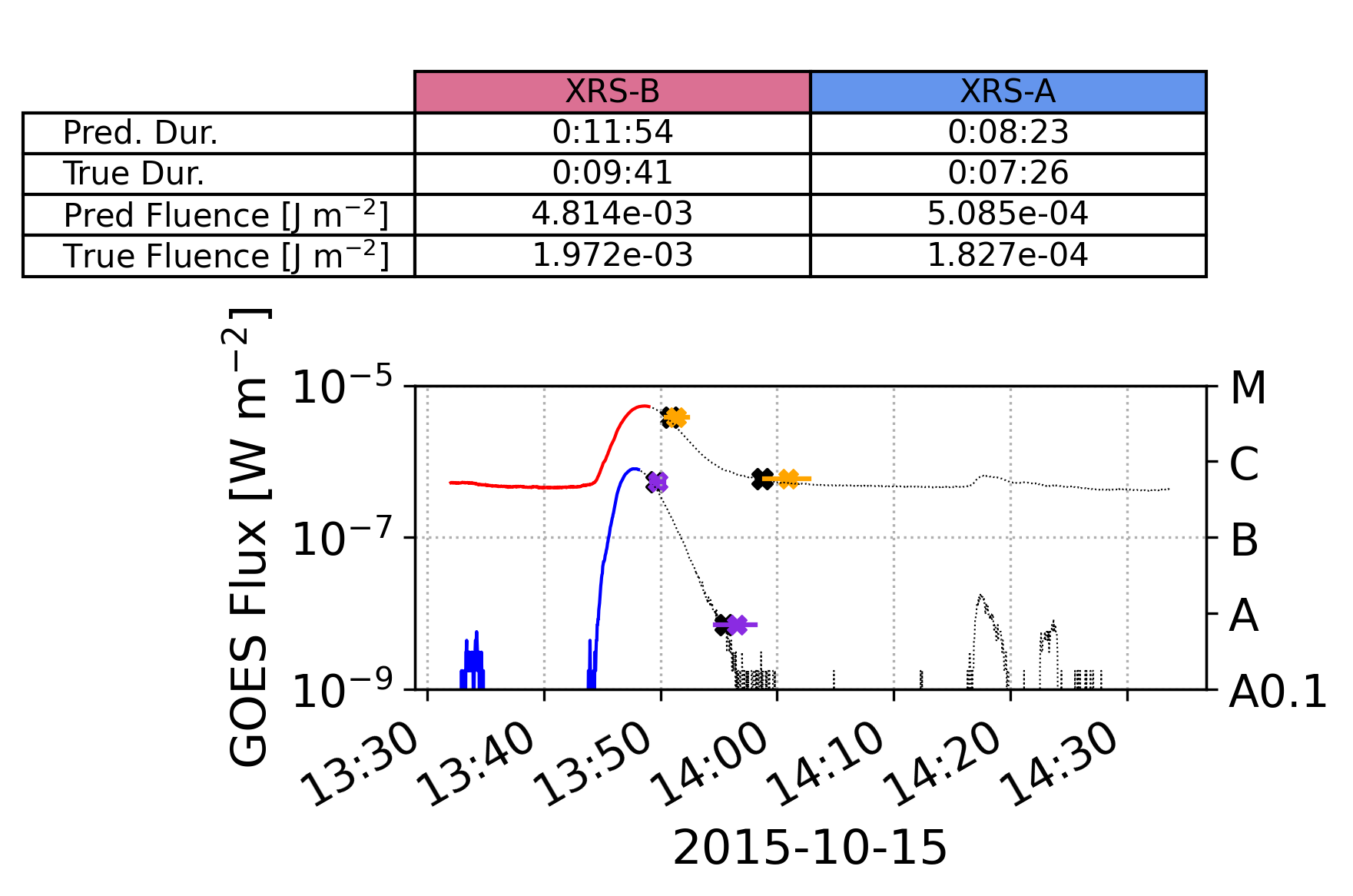}
\includegraphics[width=0.5\textwidth]{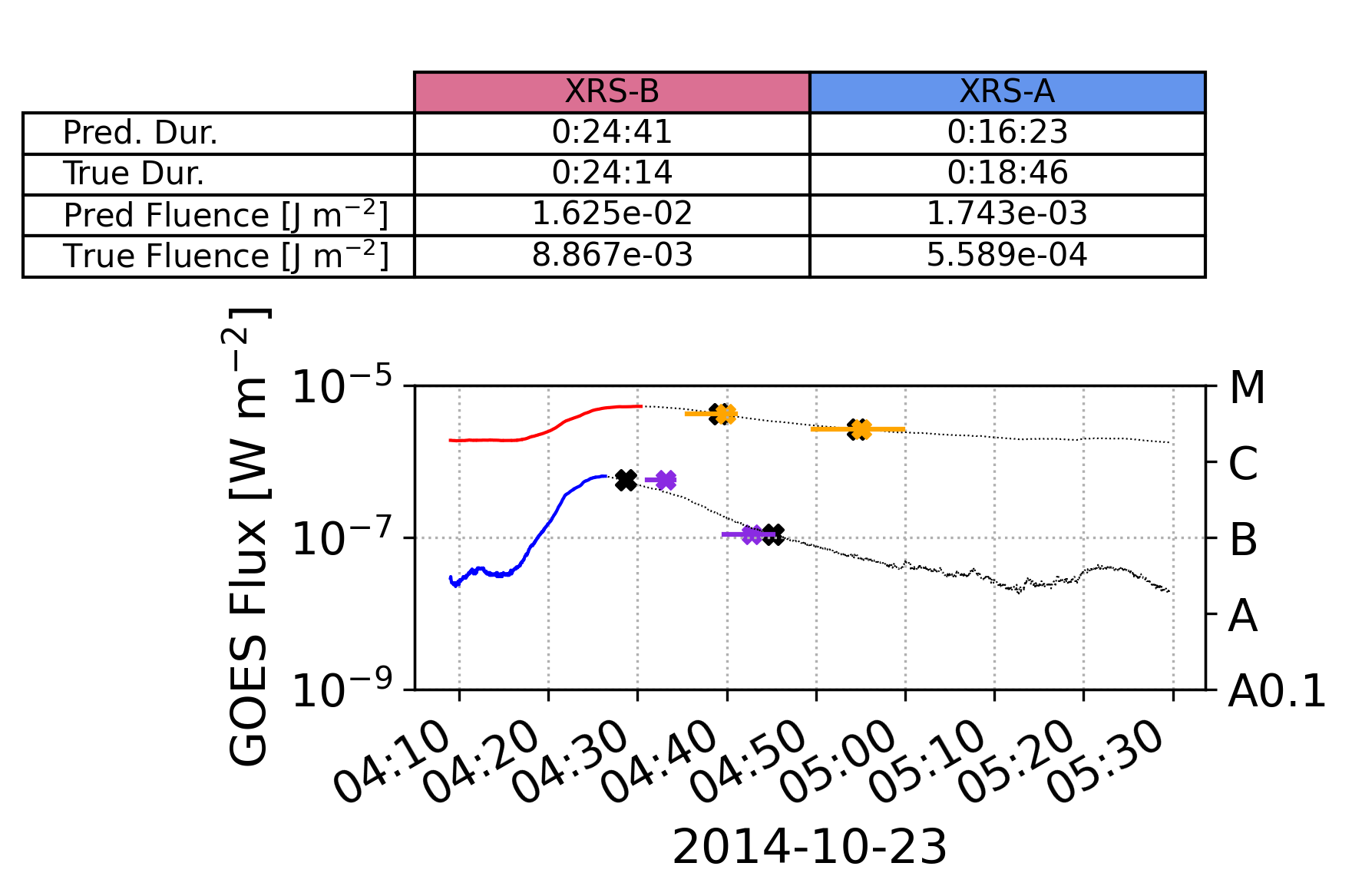}
\includegraphics[width=0.5\textwidth]{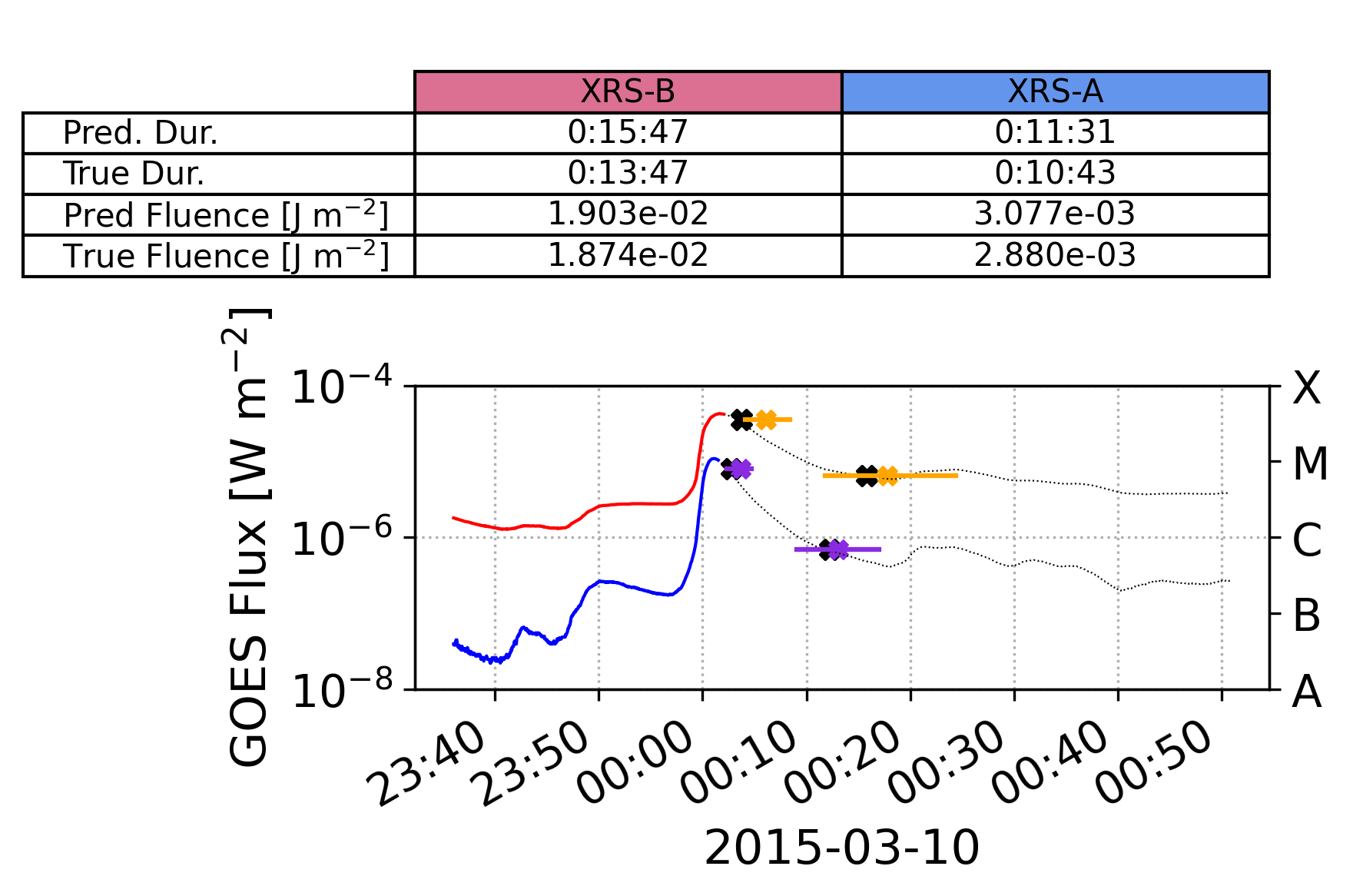}
\includegraphics[width=0.5\textwidth]{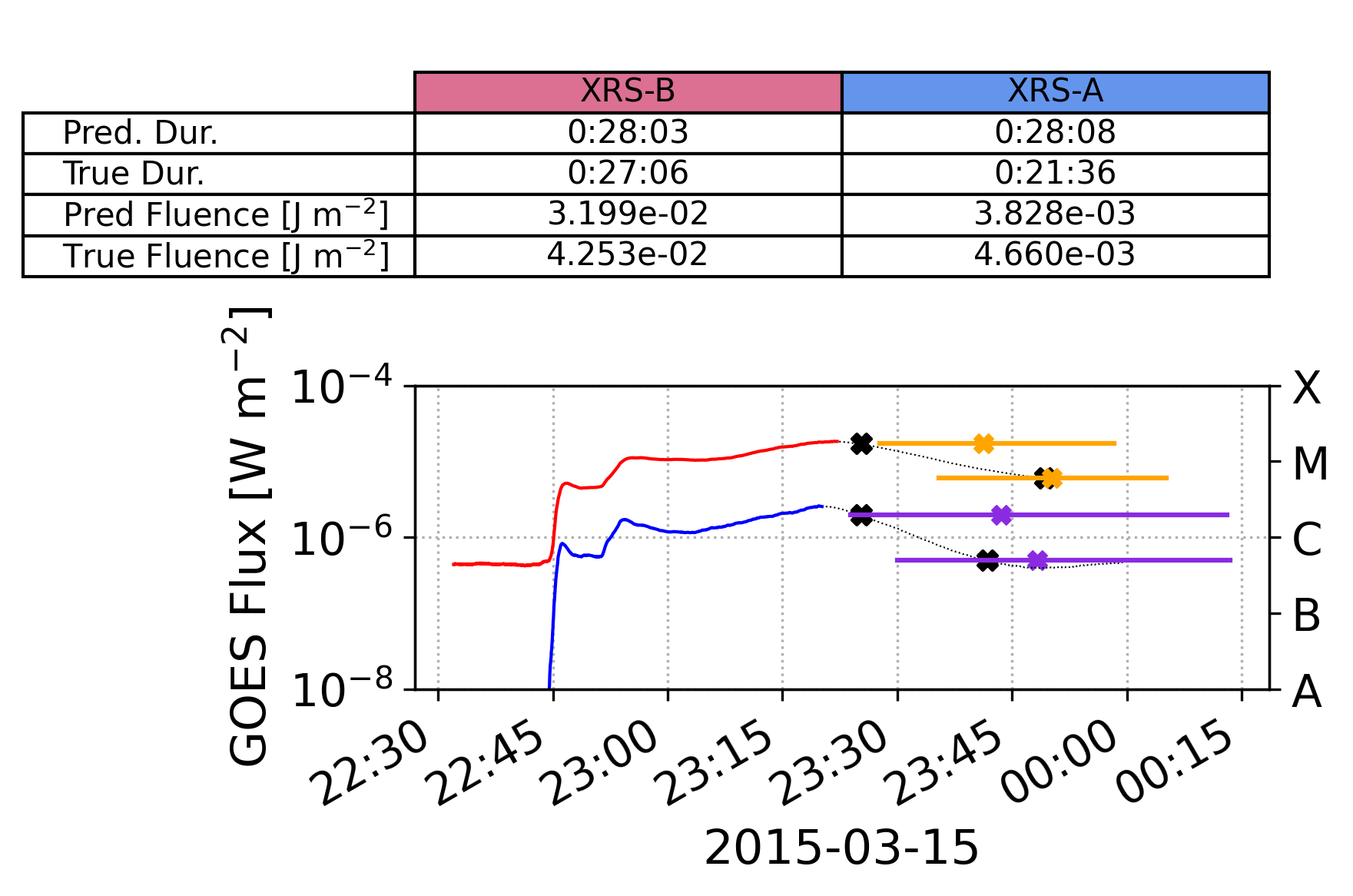}
\includegraphics[width=0.5\textwidth]{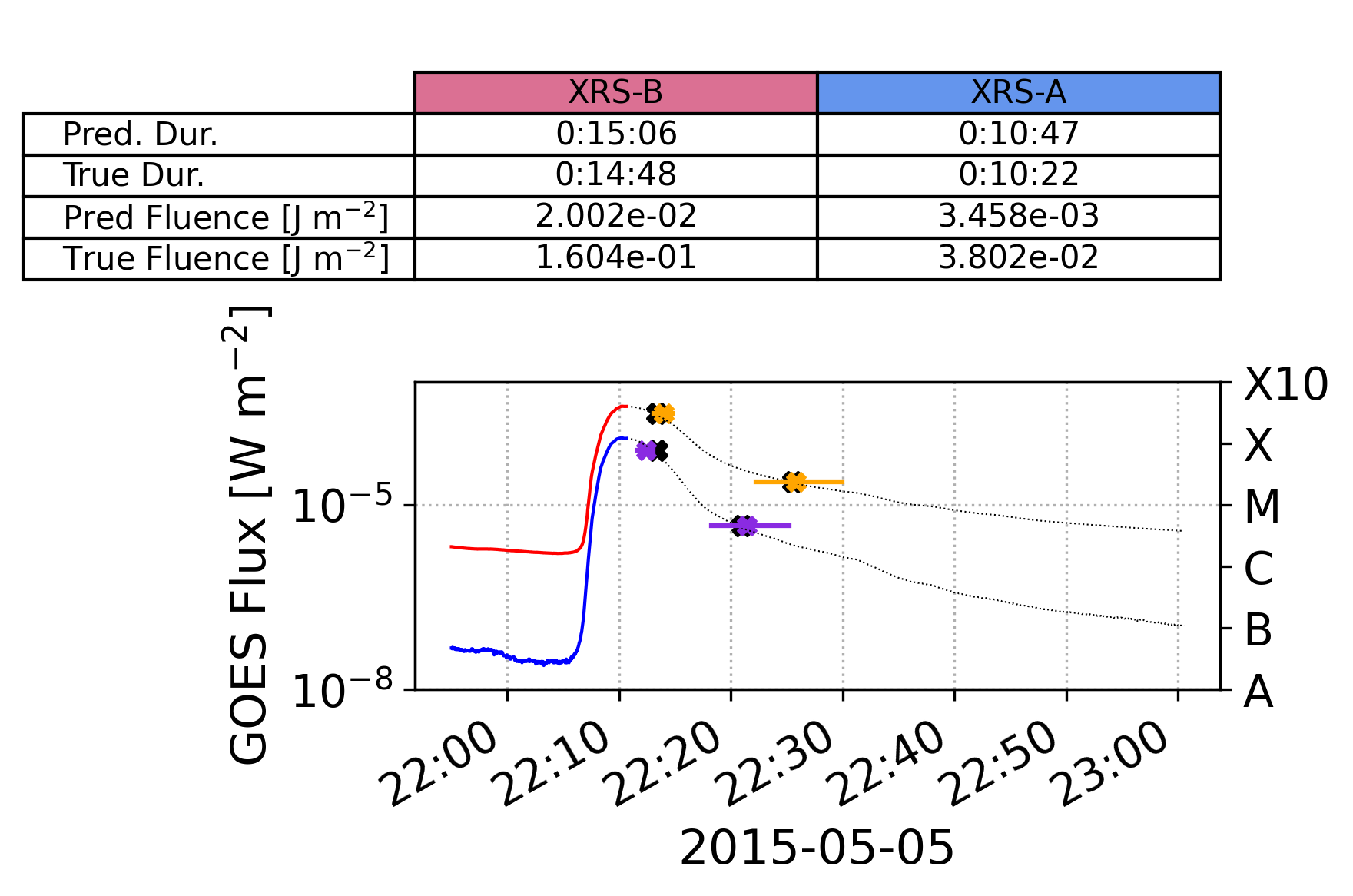}
\includegraphics[width=0.5\textwidth]{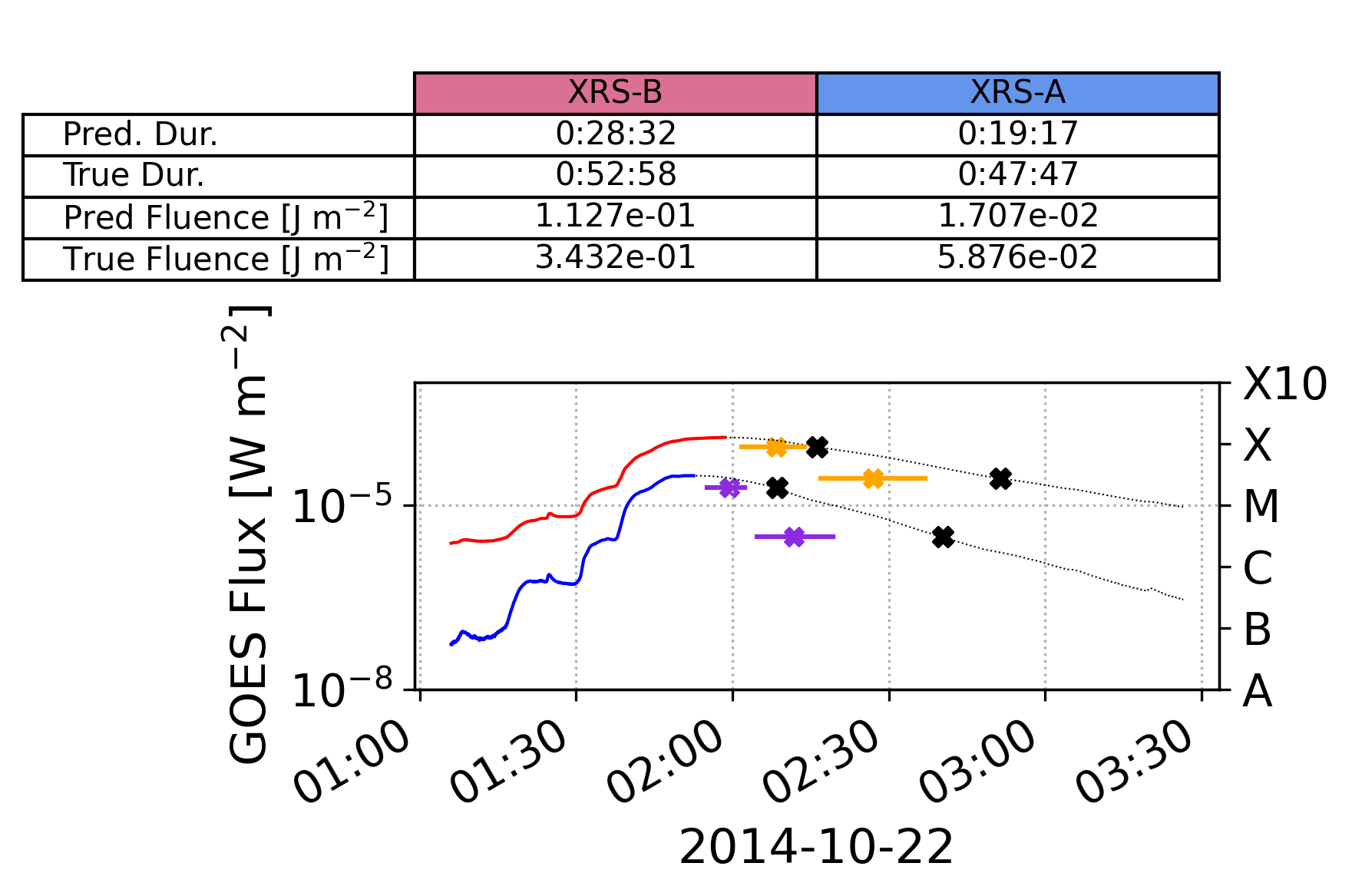}
\caption{Forecasts of ($t_{4}$ - $t_{2}$) calculated at the peak for six example flares, with relatively short (left column) and long (right column) FWHMs.  The top row shows two C-class flares, the middle row M-class, and the bottom row X-class.  The red lines are for XRS-B, and blue for XRS-A.  The orange and violet X marks are the predicted values of $t_{3}$ and $t_{4}$ in each channel, compared to the true values marked in black.  The $\pm$ 1-$\sigma$ uncertainties are marked with horizontal lines.  The black dotted lines show the true evolution of the light curves.  The method tends to predict $t_{4}$ in short-FWHM flares accurately, while it tends to underestimate $t_{4}$ in longer events.  In the table above each plot, we also annotate the true and predicted values of $t_{4}$ and the fluence $A_{4}$ for each XRS channel.  \label{fig:examples_t2}} 
\end{figure*}

{To demonstrate that these predictions are reasonable, we show six example flares in Figure \ref{fig:examples_t2} with both relatively short (left) and long (right) FWHMs, ranging in class from C-class (top), M-class (middle), to X-class (bottom).  We estimate the remaining duration, $t_{4} - t_{2}$, from the peak of each channel at time $t_{2}$.  On each plot, we show the predicted values of $t_{3}$ and $t_{4}$ with an X mark, where the orange case shows the prediction for XRS-B and violet for XRS-A, and the horizontal lines show the $\pm$ 1-$\sigma$ uncertainties.  The true values of $t_{4}$ are marked in black.  In general, the predictions of $t_{4}$ for the shorter FWHM flares are close to the true values, while those for longer FWHM flares are underestimated.  This agrees with Figure \ref{fig:predt2_t3t4}, which shows that the predictions for $t_{4}$ scale more slowly than the true values (slope $< 1$).  We note the estimated and true remaining time until $t_{4}$, as well as the predicted and true fluence values $A_{4}$.}   

\subsection{Predictions at time $t_{3}$}
\begin{figure*}
\includegraphics[width=0.5\textwidth]{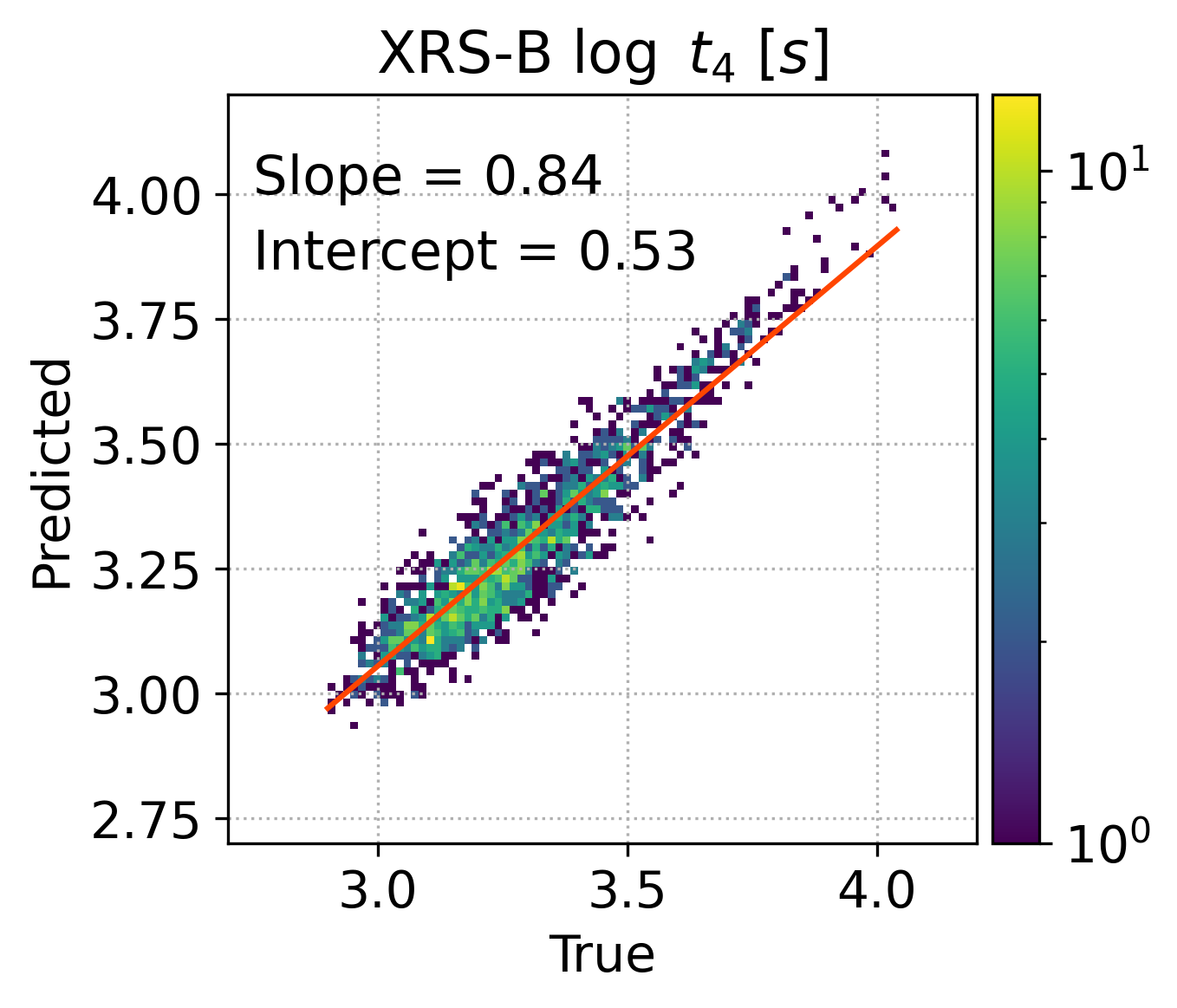}
\includegraphics[width=0.5\textwidth]{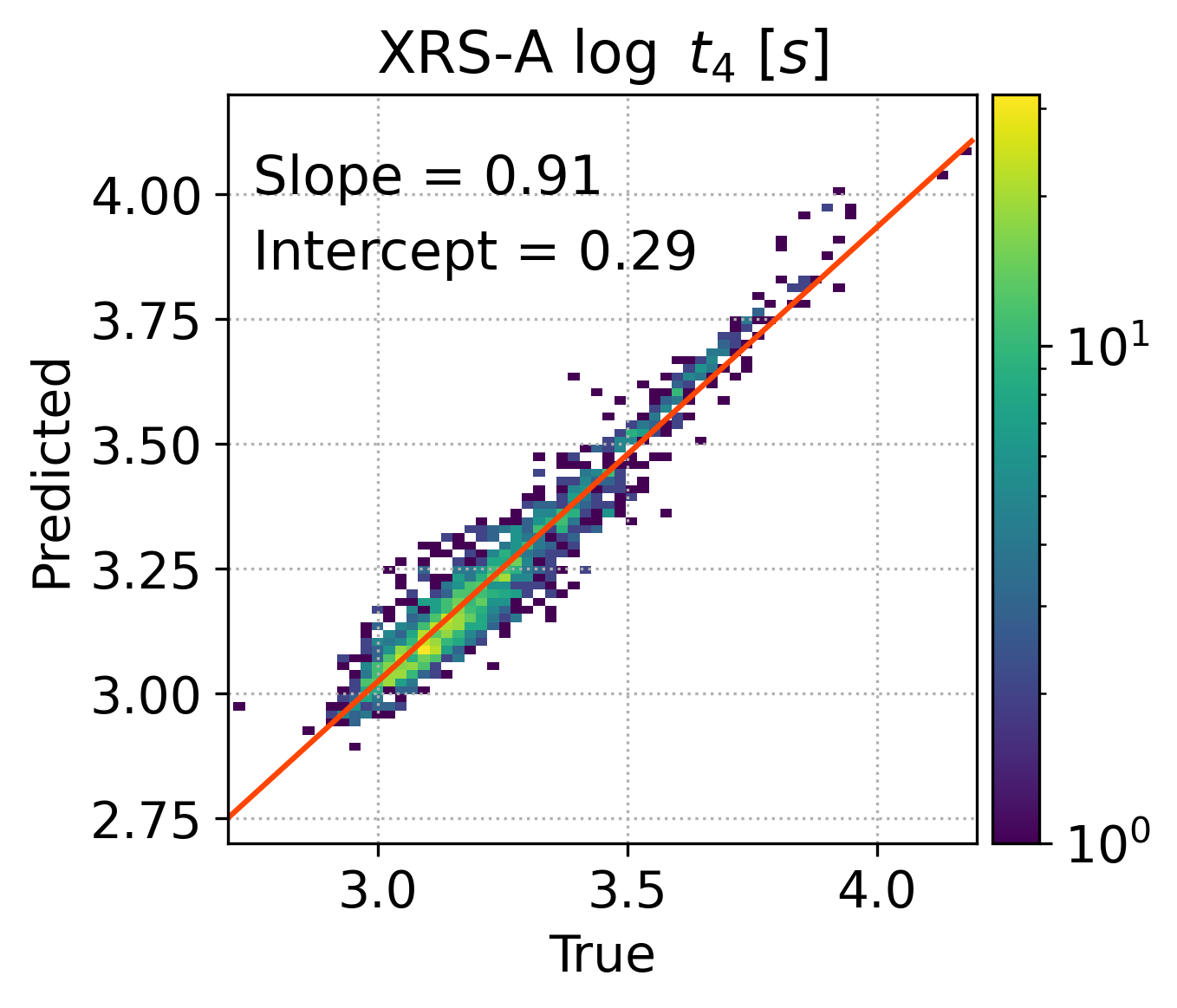}
\includegraphics[width=0.5\textwidth]{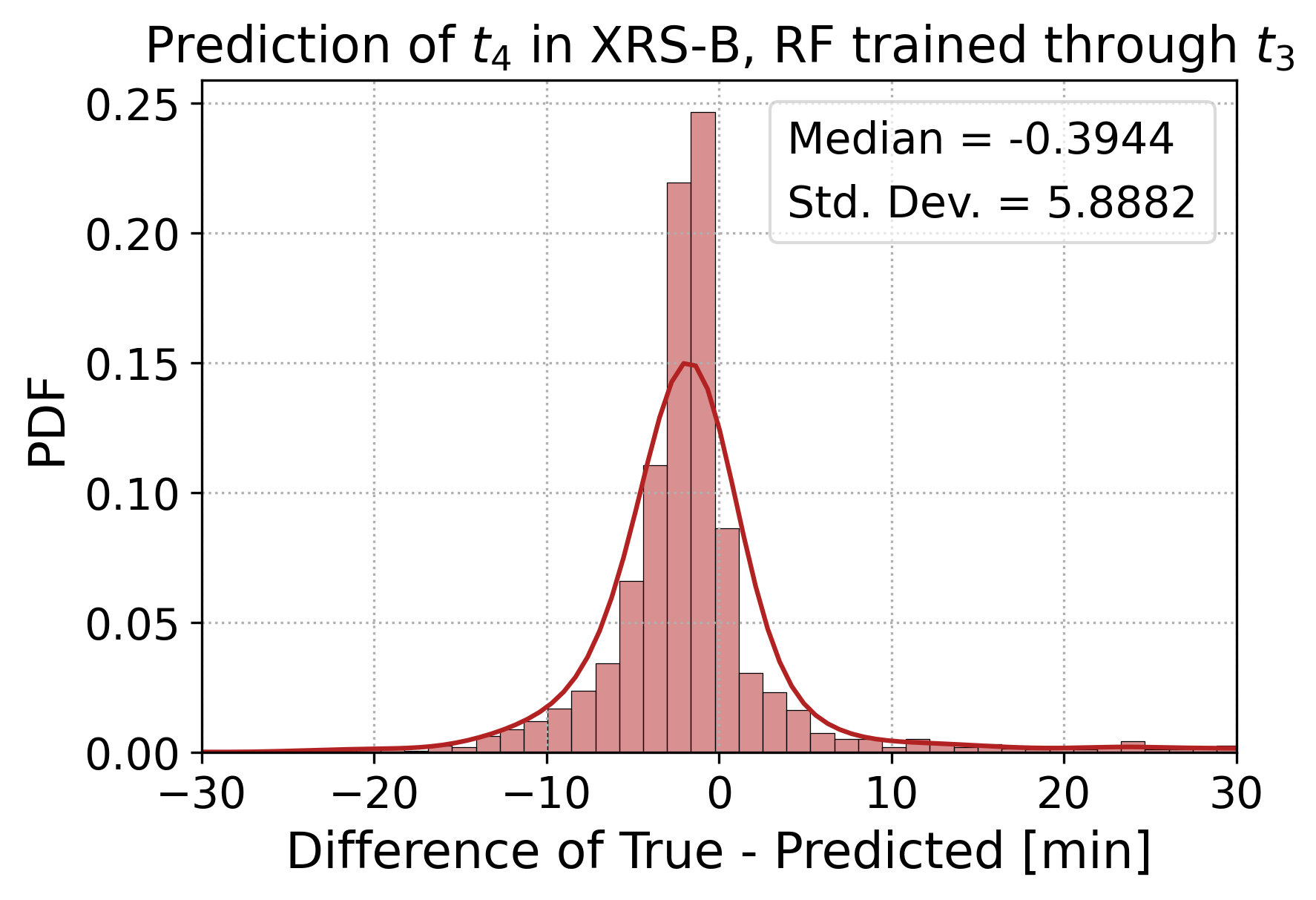}
\includegraphics[width=0.5\textwidth]{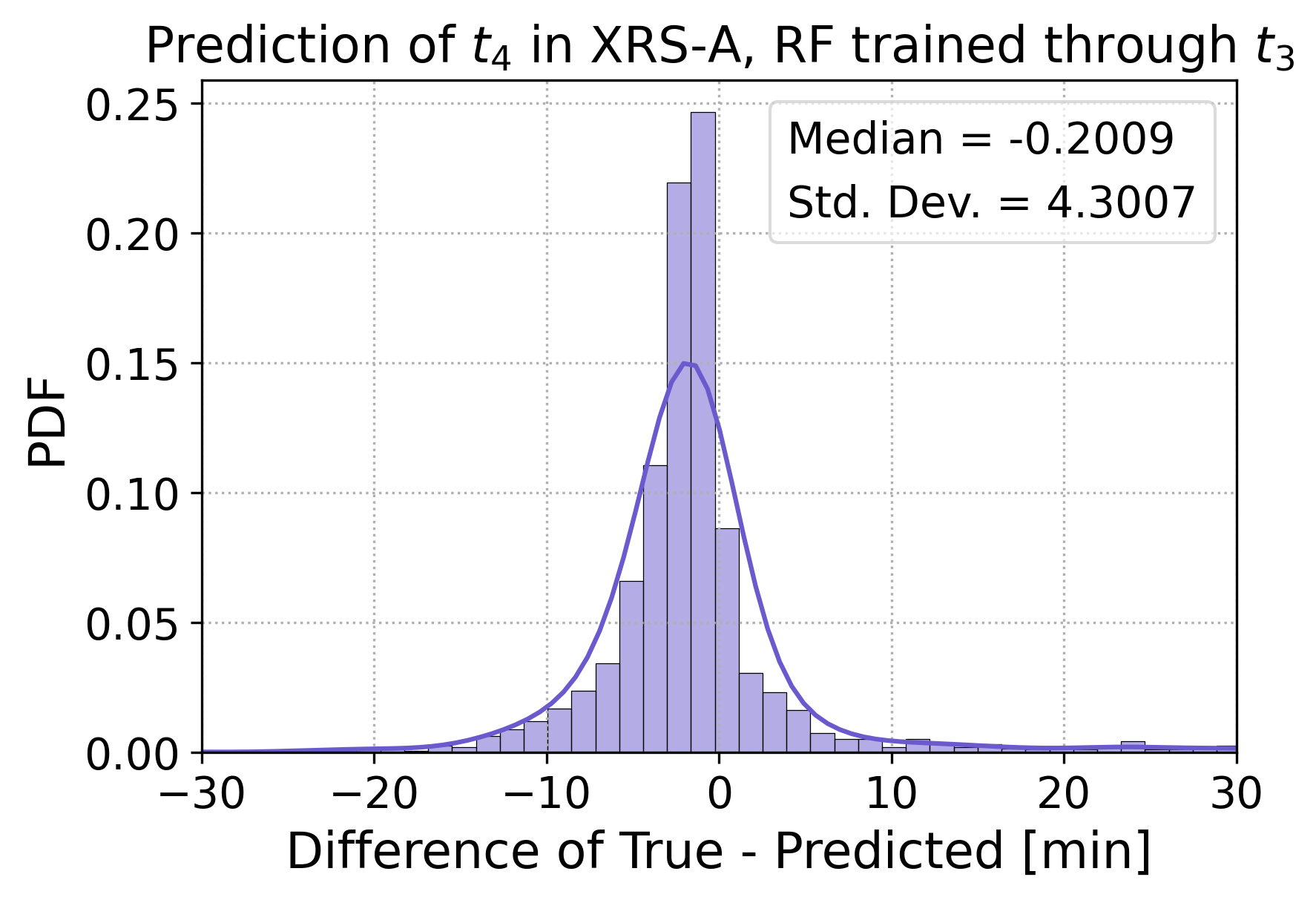}
\includegraphics[width=0.5\textwidth]{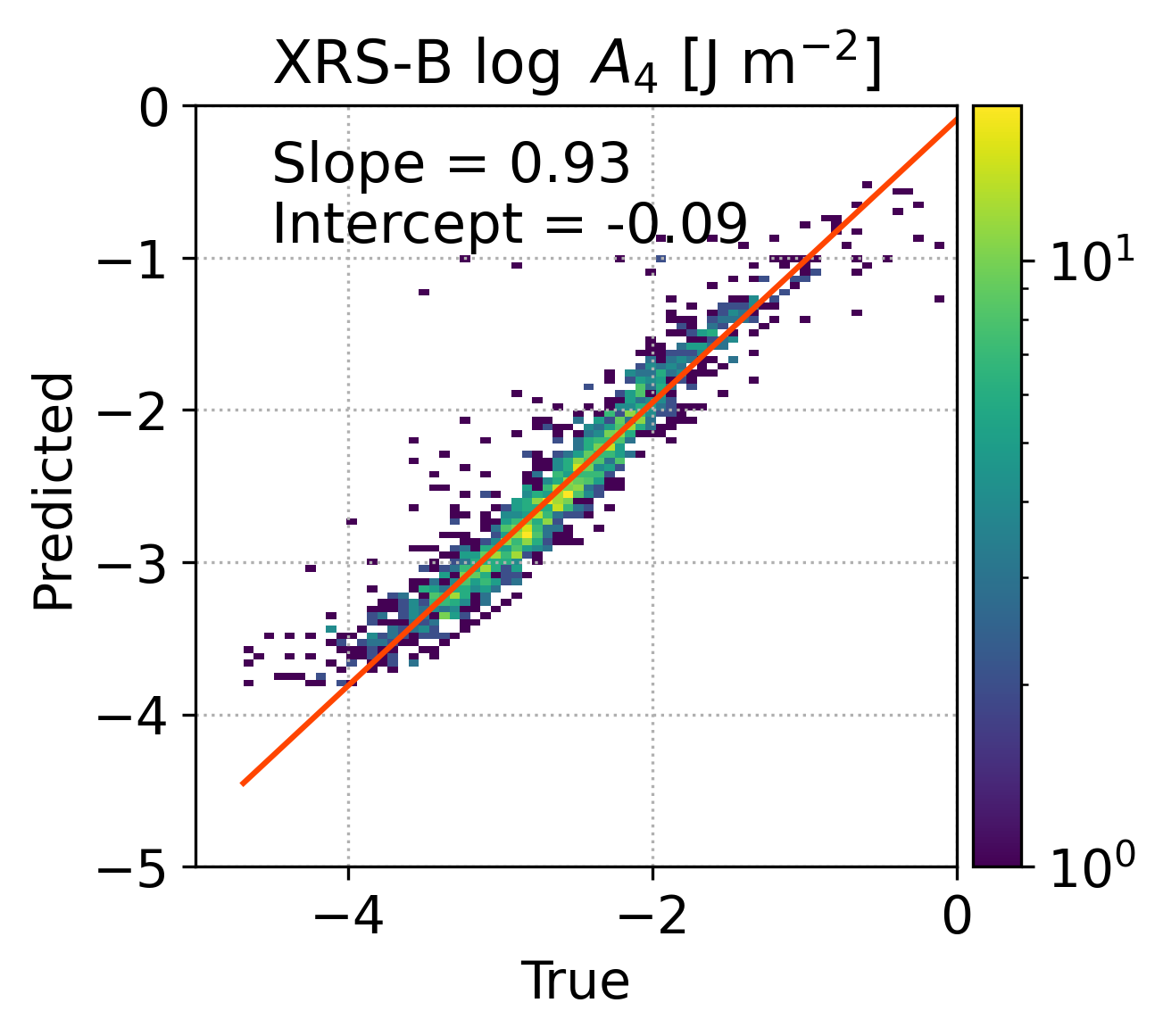}
\includegraphics[width=0.5\textwidth]{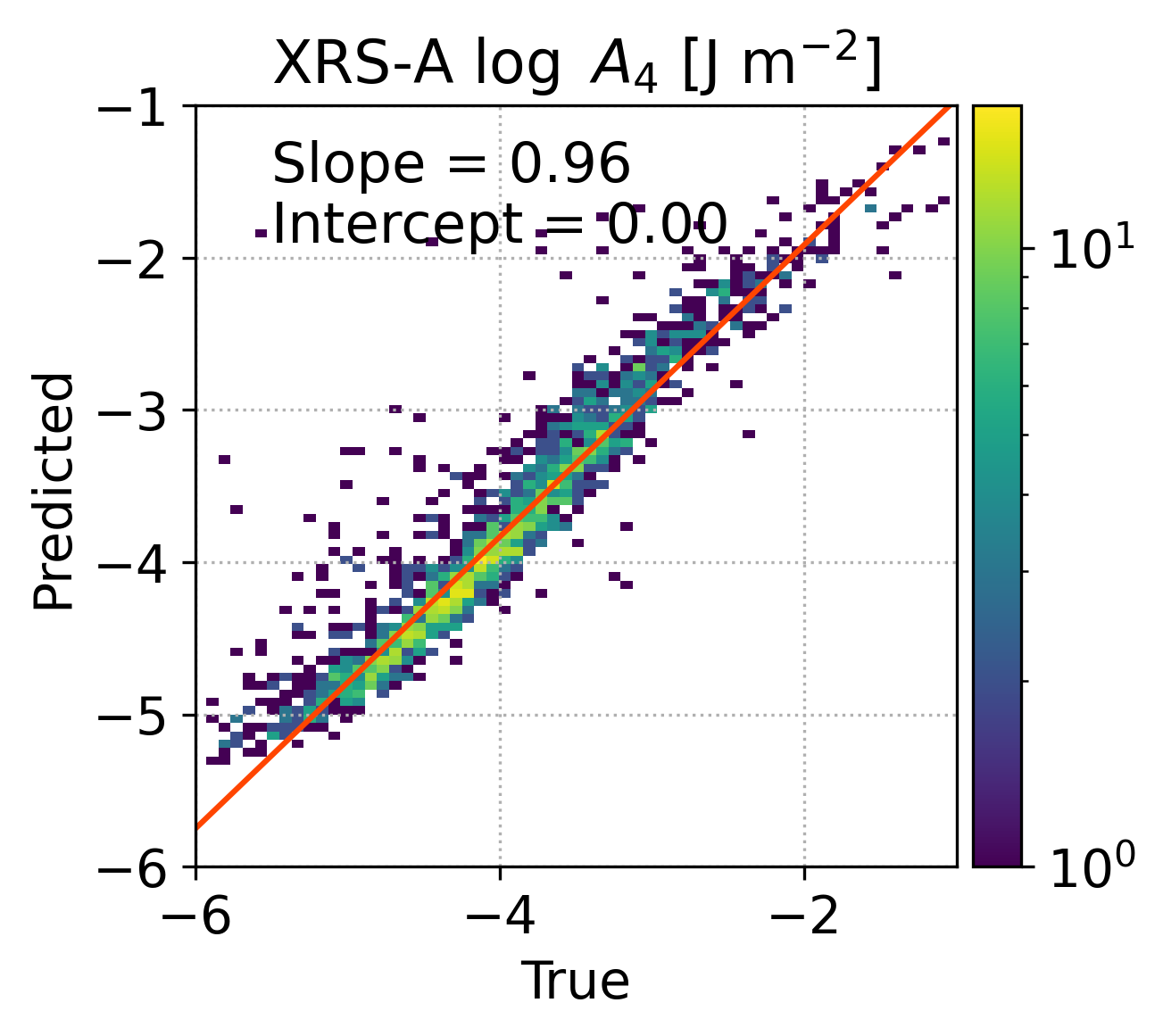}
\caption{Similar to Figure \ref{fig:predt2_t3t4}, except that the random forest has now been trained using features through time $t_{3}$ to predict the values of $t_{4}$.  The slopes are closer to 1 in this case, indicating a better fit, though still slightly underestimating $t_{4}$ in general.  This can also be seen in the PDFs of the predictions' errors, shown in the center row, indicating median errors of -0.6 and -0.2 minutes.  We also show the predicted fluences $A_{4}$, and similarly note that their predictions have improved relative to those in Figure \ref{fig:predt2_A}.}
\label{fig:predt3_t4}
\end{figure*}

We repeat the analysis, this time training the random forest using the parameters through time $t_{3}$ (the minima of the first derivatives) in order to forecast $t_{4}$ and $A_{4}$. We use the same train-test split of the data as before to ensure a valid comparison.  Figure \ref{fig:predt3_t4} shows the results of the new predictions for the two XRS channels.  In this case, the slopes are approximately $0.9$, improving upon the estimates in Figure \ref{fig:predt2_t3t4}.  The scatter is similarly reduced.  In general, the method still slightly underestimates $t_{4}$ in many events.  The error is on average about -0.4 and -0.2 minutes in XRS-B and A, respectively, showing a noticeable improvement over the previous set of predictions.  Finally, the predicted fluences $A_{4}$ are similarly improved (compared with Figure \ref{fig:predt2_A}), with a slope closer to 1 and reduced scatter.   

\begin{figure*}
\includegraphics[width=0.5\textwidth]{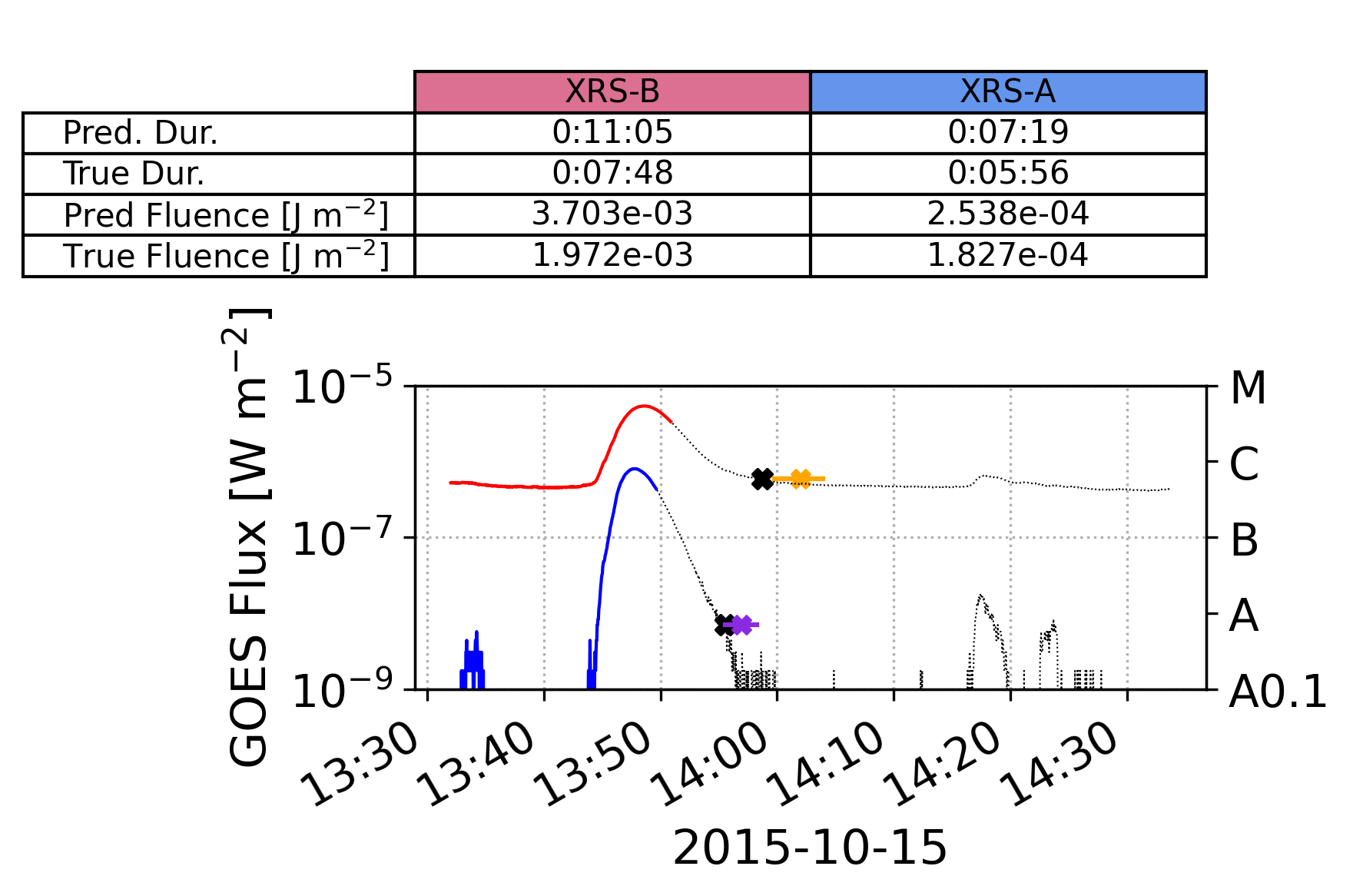}
\includegraphics[width=0.5\textwidth]{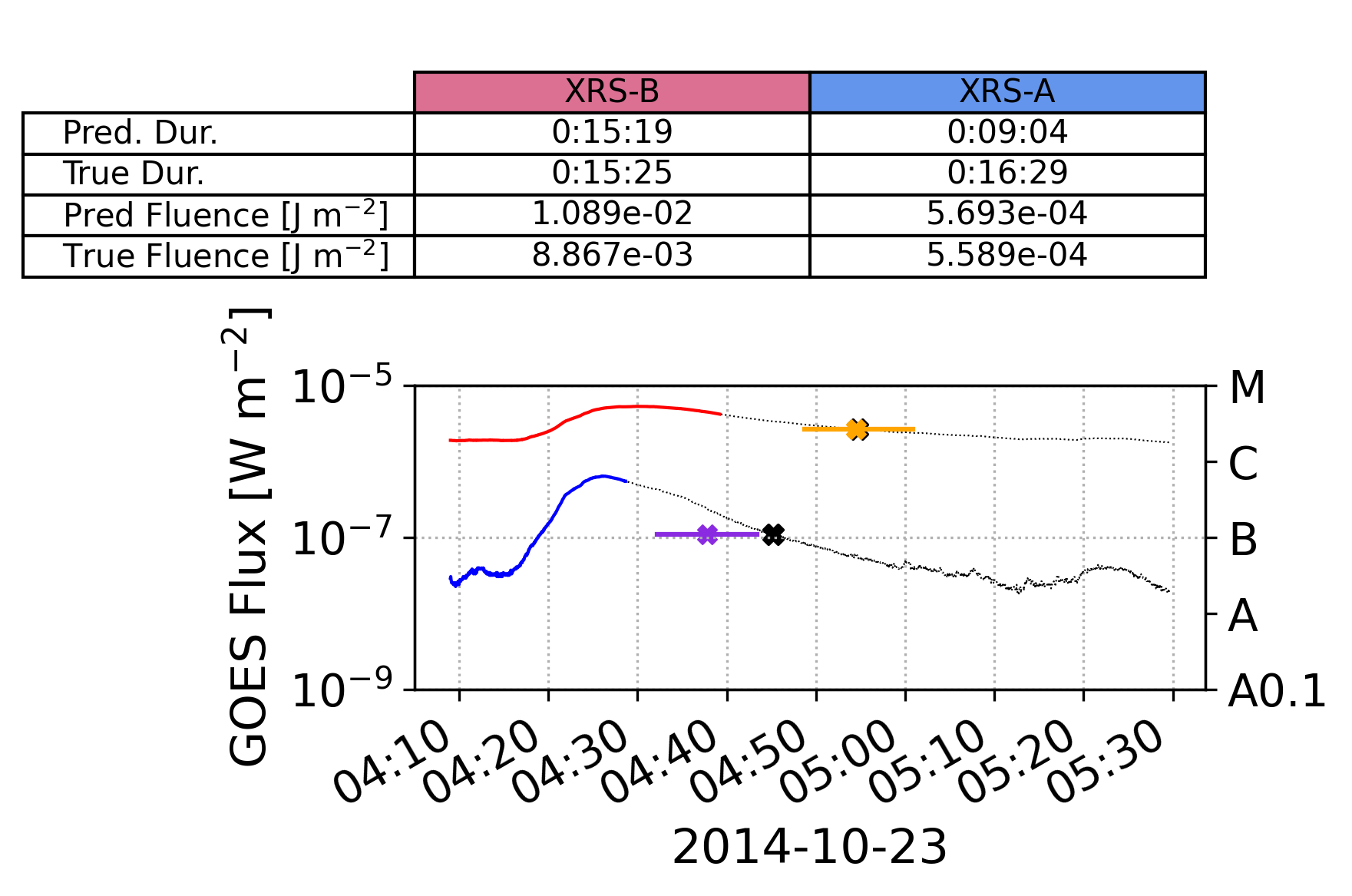}
\includegraphics[width=0.5\textwidth]{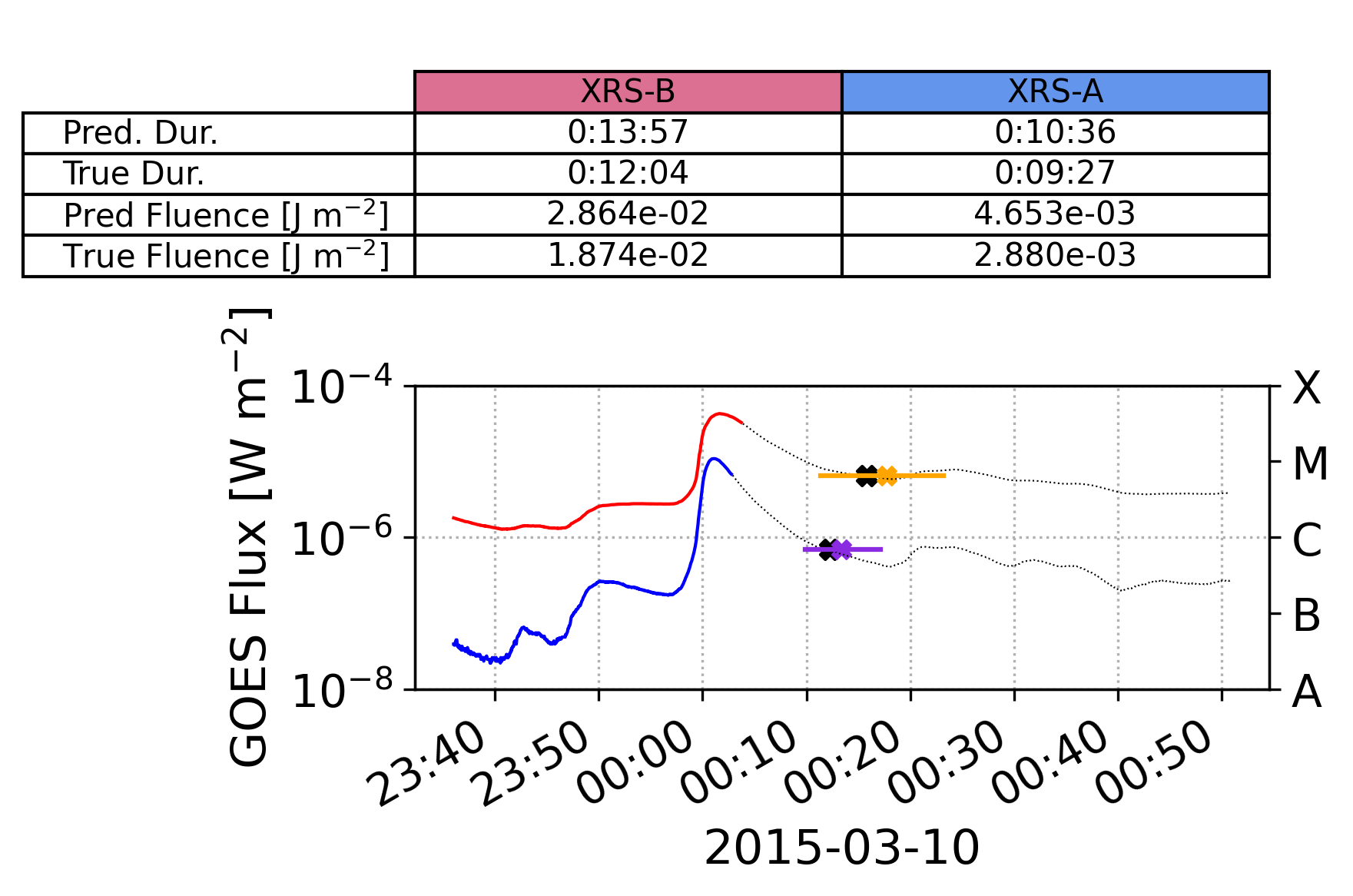}
\includegraphics[width=0.5\textwidth]{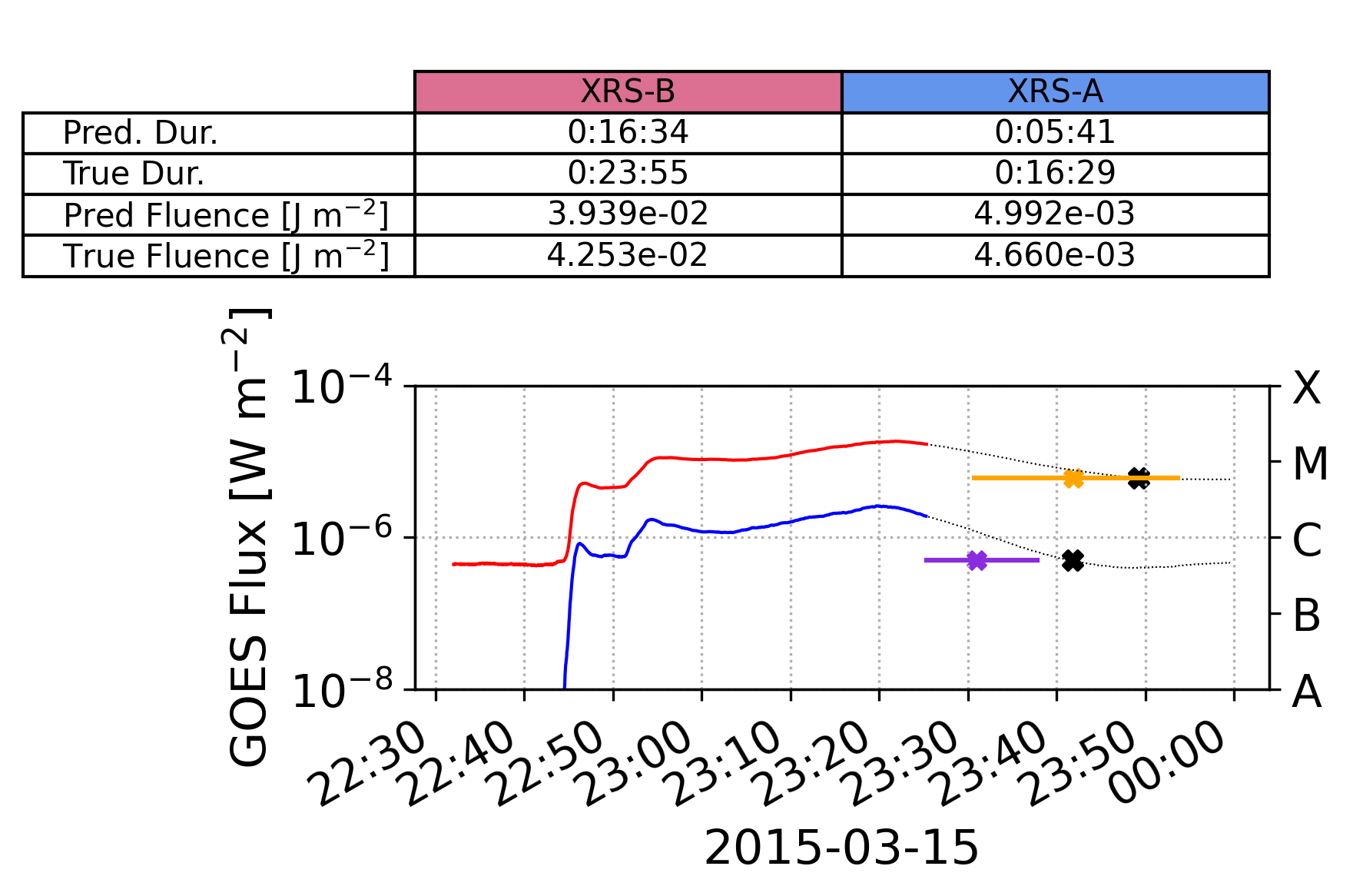}
\includegraphics[width=0.5\textwidth]{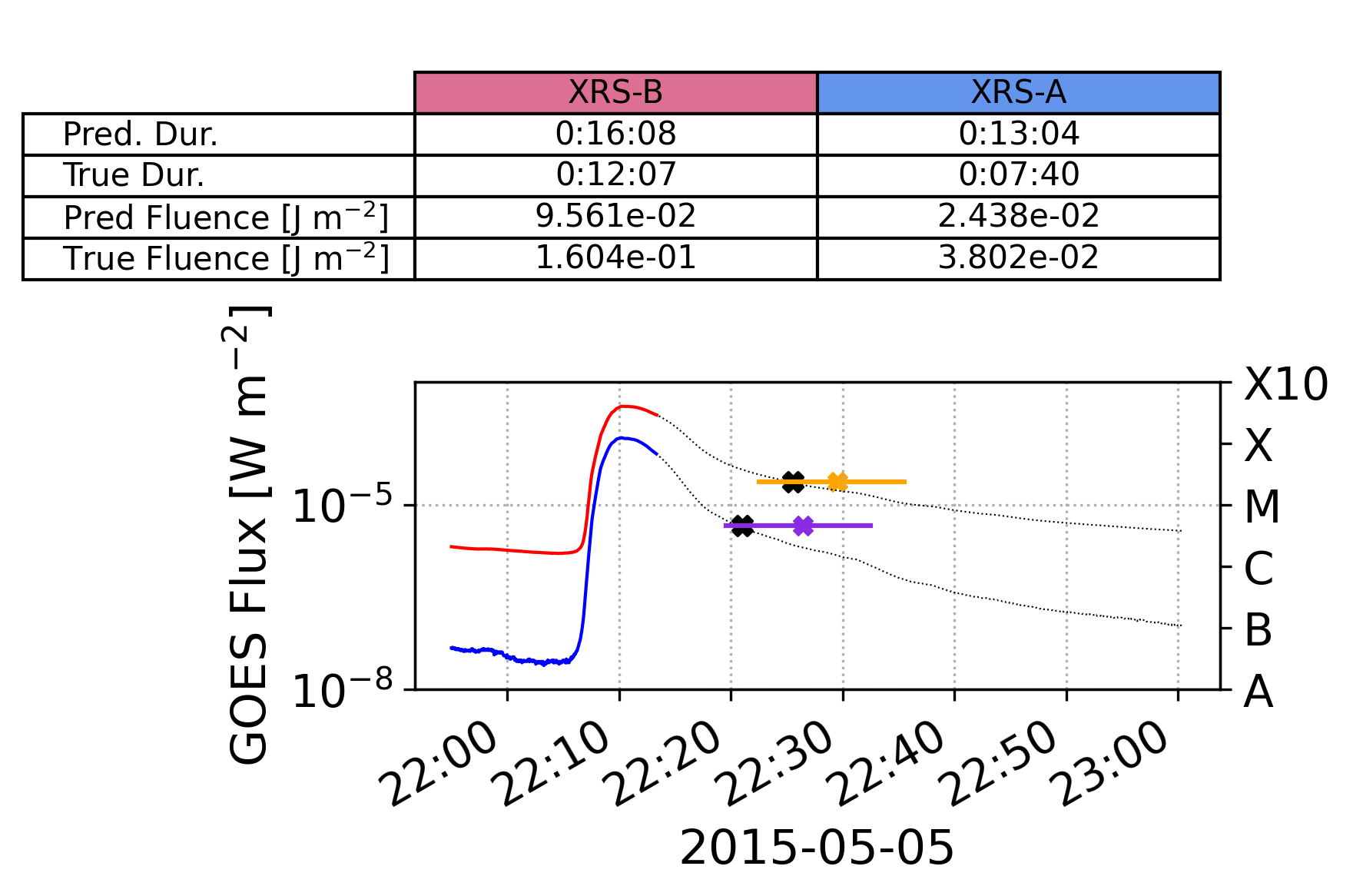}
\includegraphics[width=0.5\textwidth]{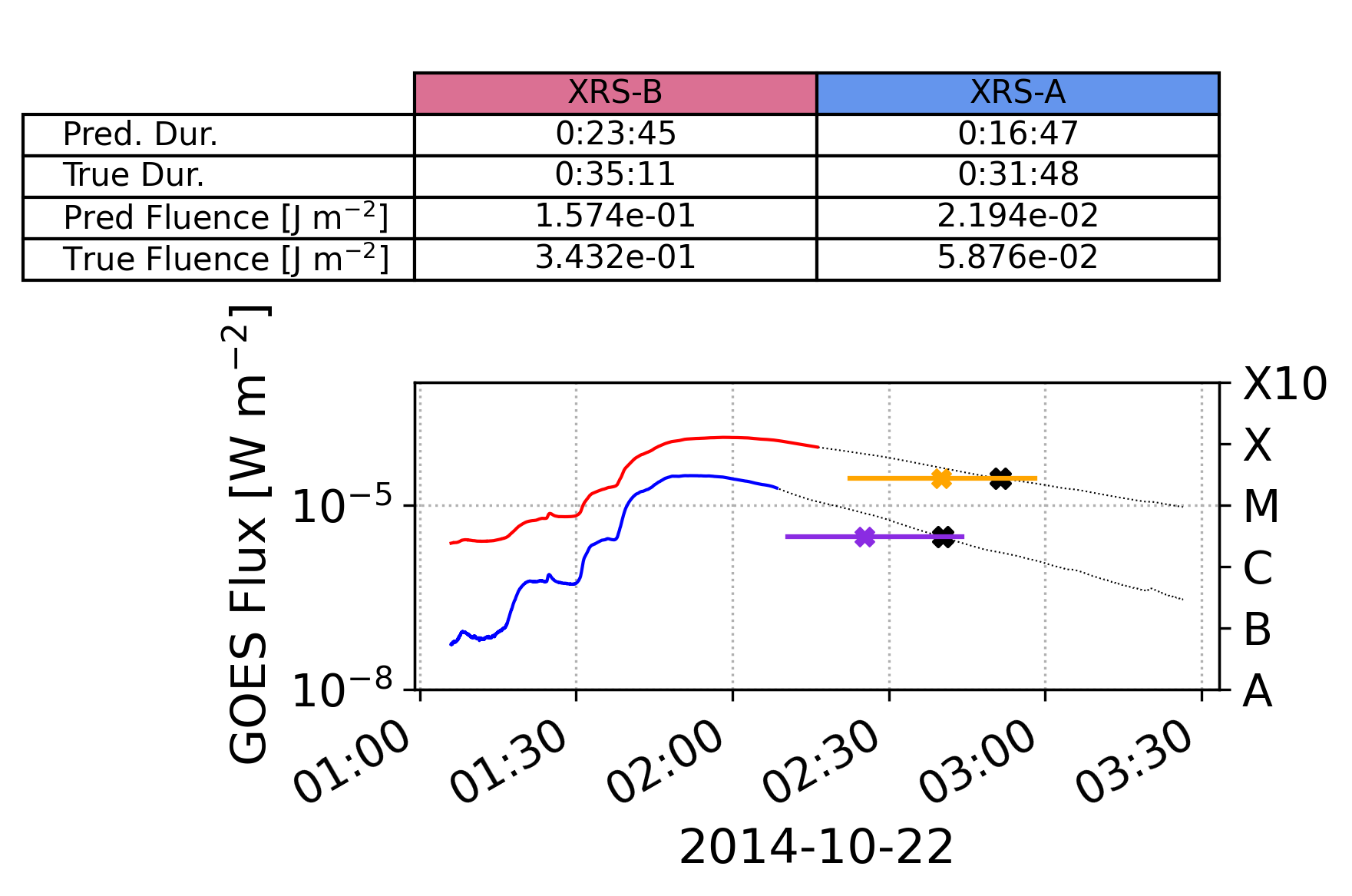}
\caption{Forecasts of ($t_{4}$ - $t_{3}$) and $A_{4}$ similar to Figure \ref{fig:examples_t2}, showing the predictions beginning at time $t_{3}$ (the minimum of the first derivative) for the same six events.  \label{fig:examples_t3}} 
\end{figure*}

In Figure \ref{fig:examples_t3}, we show the predictions for the same six flares presented previously, including C, M, and X class events with both short and long FWHMs.  We see that the predictions are now slightly improved when compared to their true values, although the estimates of $t_{4}$, particularly for the longer events, are still somewhat short.  For example, the long-FWHM X-flare, at bottom right, is still underestimated.  This suggests that as the flare proceeds from times $t_{2}$ (the peak) to $t_{3}$ (the cooling phase), we can improve our estimate of the approximate end of the flare $t_{4}$.  The predicted fluences have similarly improved compared to their estimates in Figure \ref{fig:examples_t2}.

\subsection{Prediction Skill}
\texttt{scikit-learn} also calculates the feature importance, or which features are most important in determining the predictions.  For regression problems, this is quantified by the reduction in variance for each feature, and then the features are ranked accordingly.  For predictions of $t_{4}$ made at time $t_{2}$, the most important features are, respectively, $t_{2}$, $t_{0}$, $A_{2}$, $A_{1}$, $t_{1}$.  The flux levels and derivative magnitudes are, perhaps surprisingly, almost negligible in their impact on the predictions.

\begin{figure*}
\includegraphics[width=0.5\textwidth]{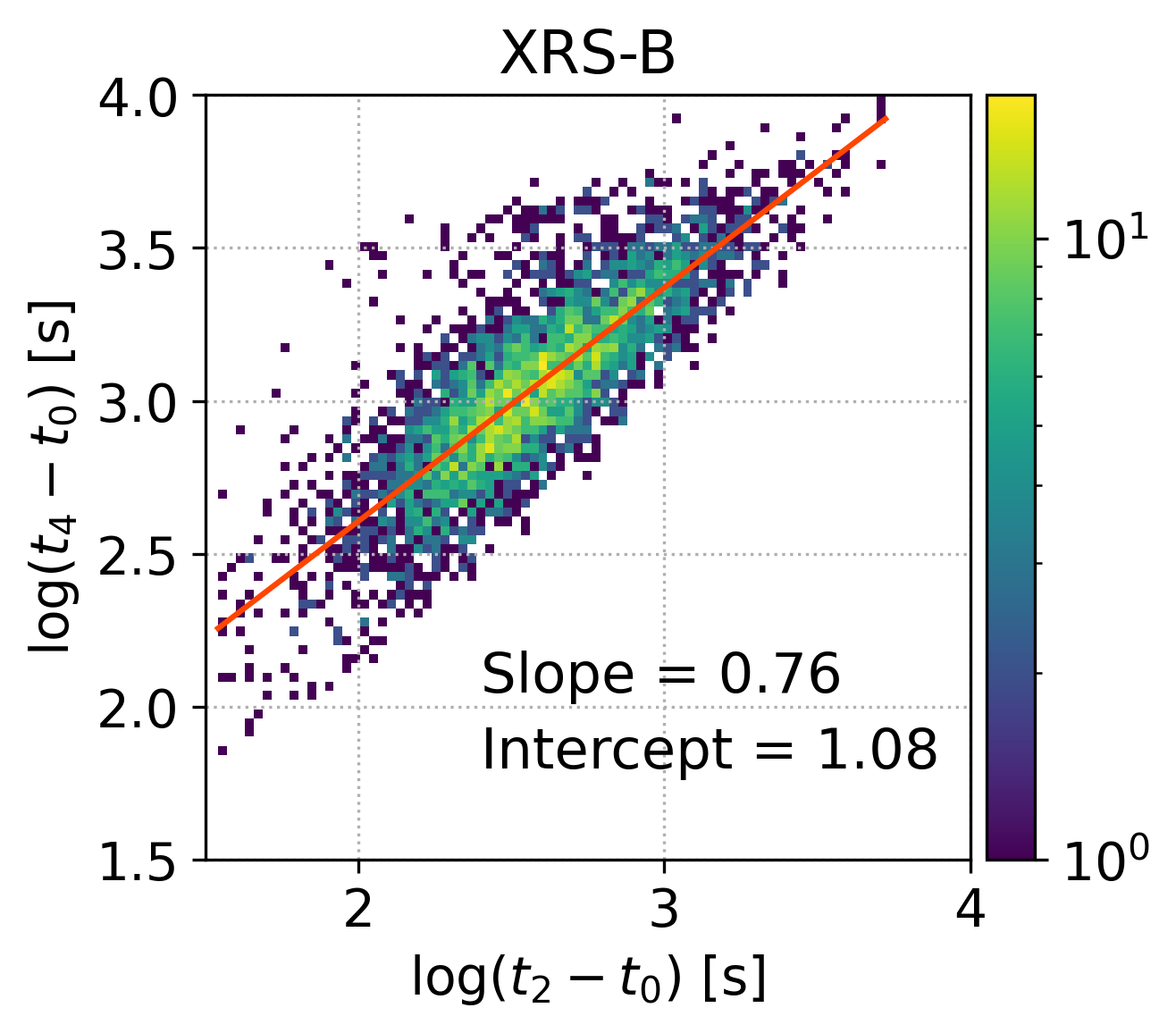}
\includegraphics[width=0.5\textwidth]{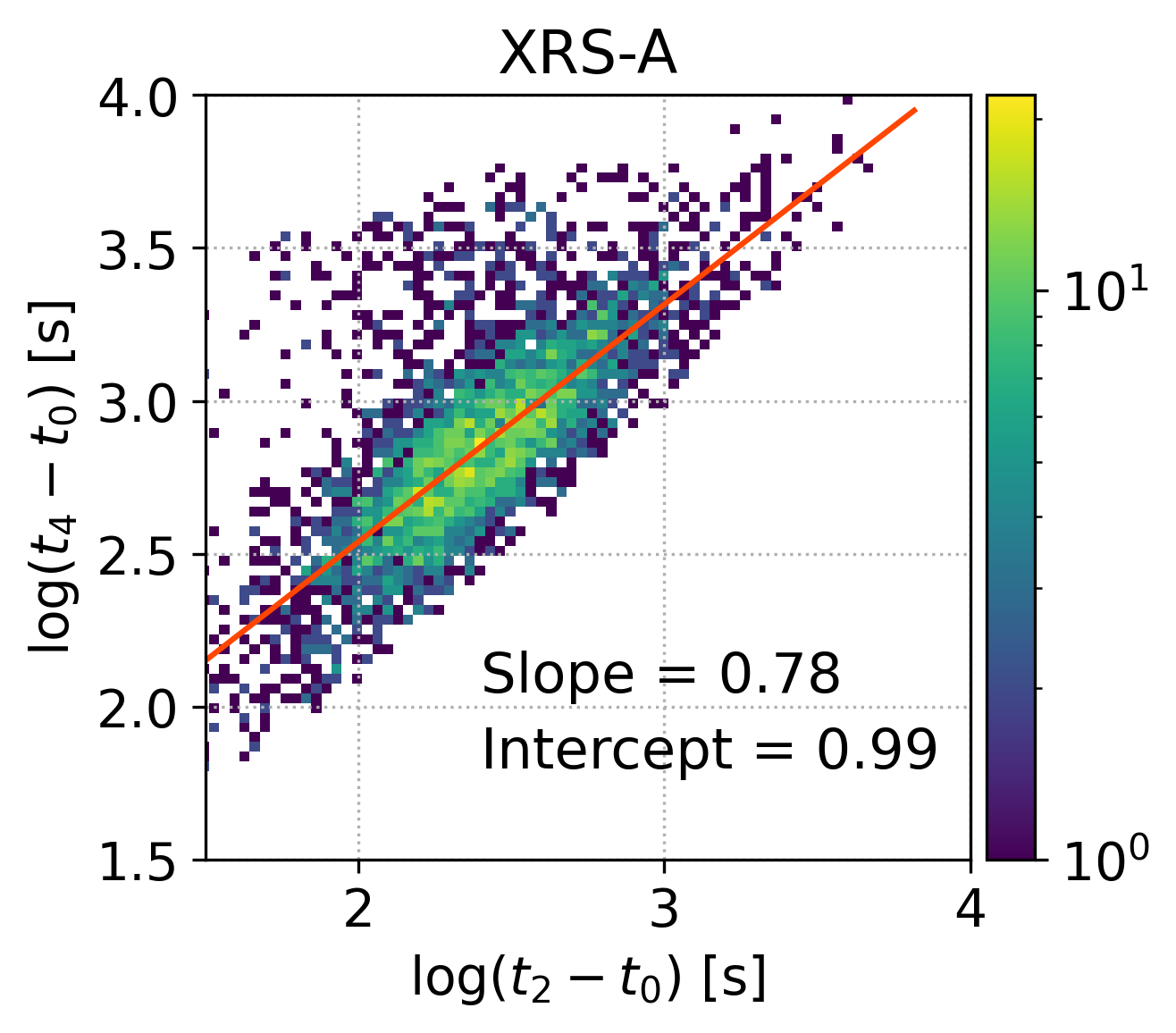}
\caption{2D histograms comparing the true values of $t_{2}$ and $t_{4}$ in the 1--8\,\AA (left) and 0.5--4\,\AA (right) channels for the entire data set.  The distributions have been fit with a Theil-Sen estimator (orange lines), for which the slope and y-intercepts are indicated.  We use this linear regression model to calculate a reference forecast which we show is outperformed by the random forest method.}
\label{fig:t2t4}
\end{figure*}
How does this prediction fare?  We calculate a skill score by comparing to a simple linear regression model.  Since $t_{2}$ is the most important feature in predicting $t_{4}$, we use it alone to give a simple prediction.  In Figure \ref{fig:t2t4}, we show scatter plots of $t_{2}$ versus $t_{4}$ for the whole data set for each GOES channel.  We also show a Theil-Sen linear regression to the data.  With these fits, we then calculate a crude estimate of the final time $t_{4,\mathrm{lin}} = m\, t_{2} + b$, for $m$ the slope and $b$ the y-intercept.  We then define a skill score as:
\begin{equation}
    \mathrm{Skill\, score} = 1 - \frac{MSE_{rf}}{MSE_{lin}}
\end{equation}
\noindent where MSE$_{rf}$ is the mean squared error from the random forest prediction for $t_{4}$ and MSE$_{lin}$ is the mean squared error from the linear regression prediction.  A skill score of exactly 1 corresponds to a perfect forecast for all events, while a skill score of 0 or less indicates that the random forest regressor was less accurate in predicting flare duration than the linear regressor.  We use only the test data to calculate the MSE to avoid biasing the skill score.  For the train-test splits and particular random forests in Section \ref{sec:data}, we find a skill score of 0.0146 in XRS-B and 0.080 in XRS-A.  For the forecasts through time $t_{3}$, these skill scores improve to 0.699 and 0.890, respectively.  That is, the random forest marginally outperforms a linear regression when the forecast occurs at time $t_{2}$ (the peak flux in each channel), and strongly outperforms the linear regression when forecasted from time $t_{3}$ (the minima of the first derivatives).

\begin{figure*}
\includegraphics[width=0.5\textwidth]{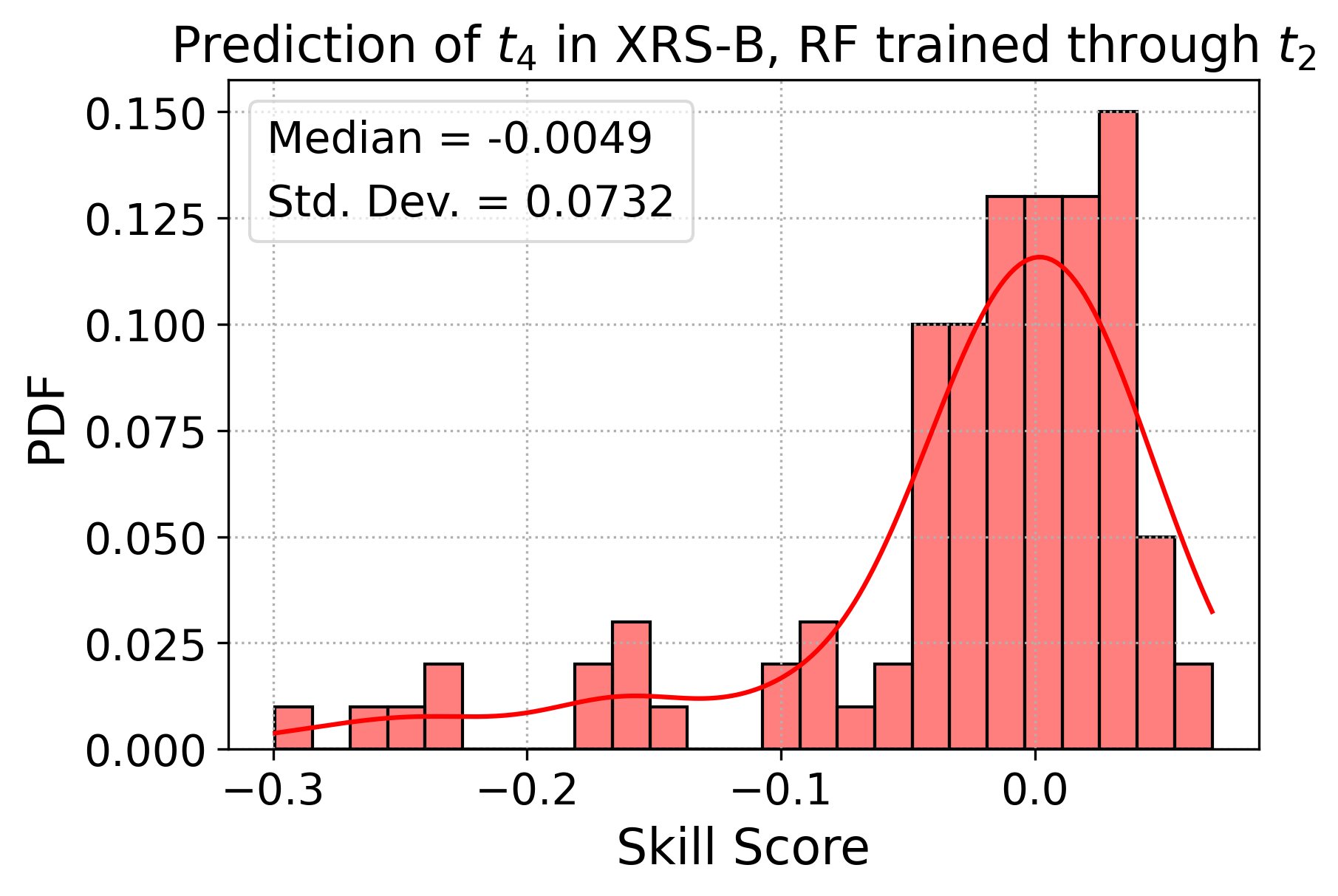}
\includegraphics[width=0.5\textwidth]{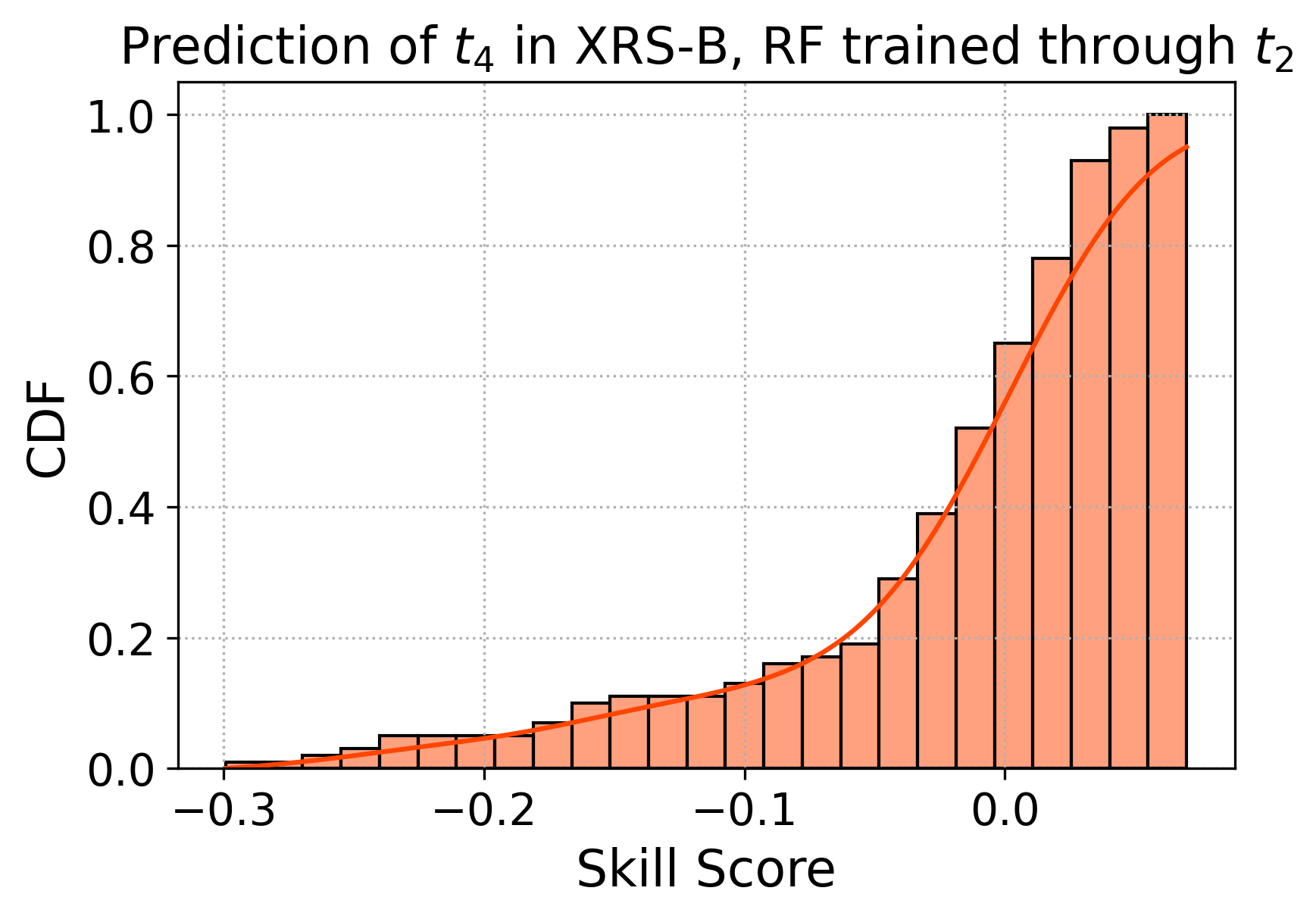}
\includegraphics[width=0.5\textwidth]{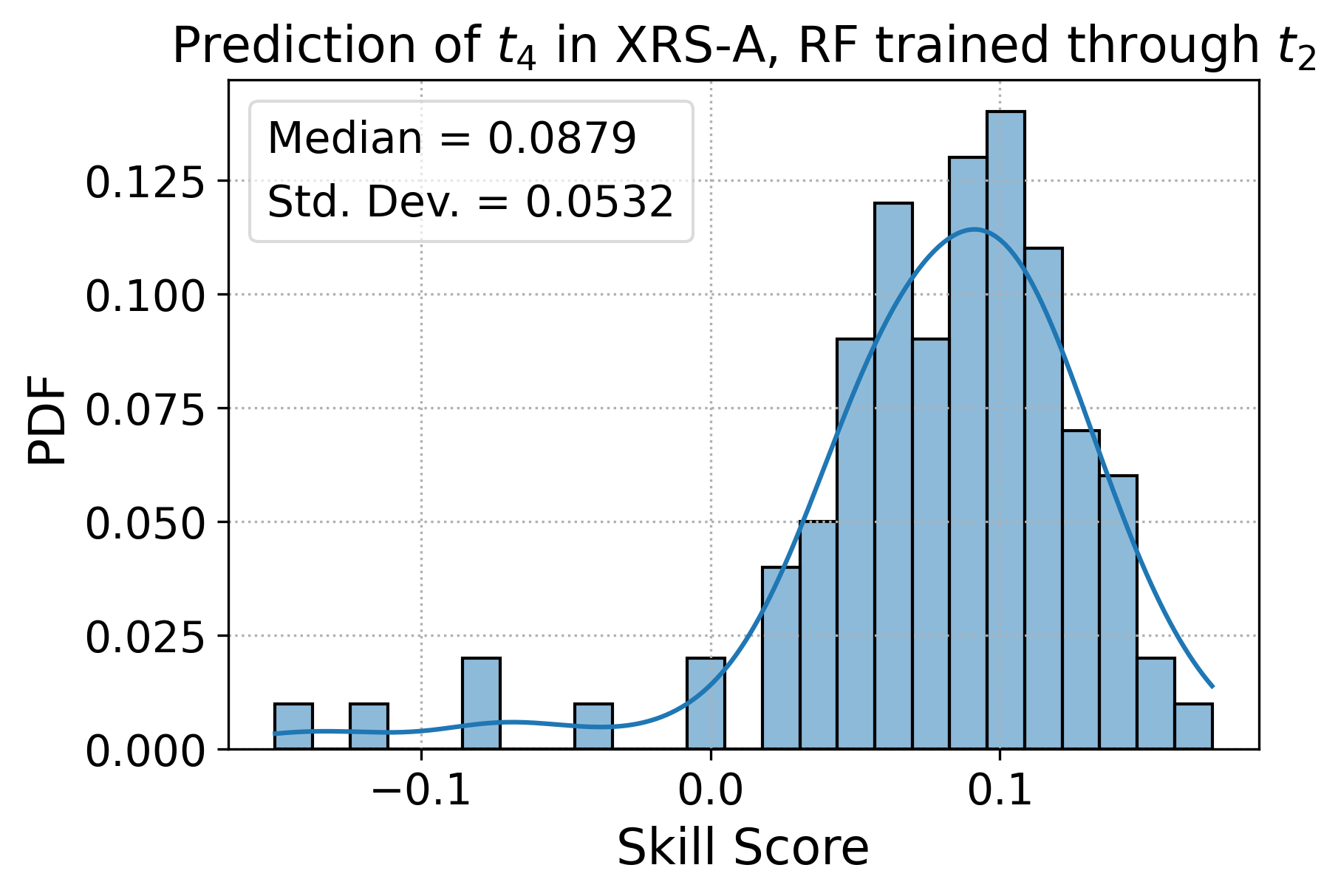}
\includegraphics[width=0.5\textwidth]{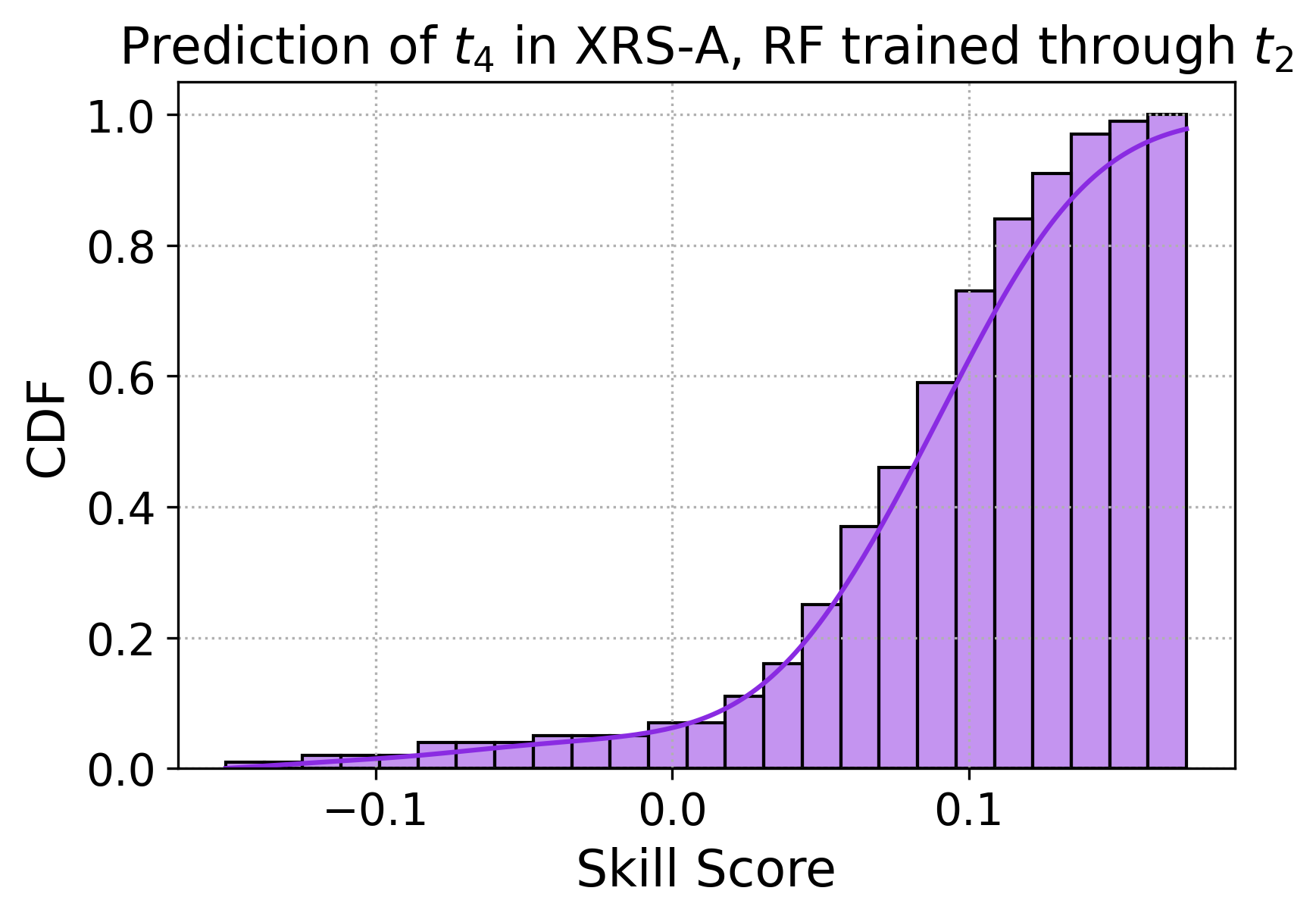}
\includegraphics[width=0.5\textwidth]{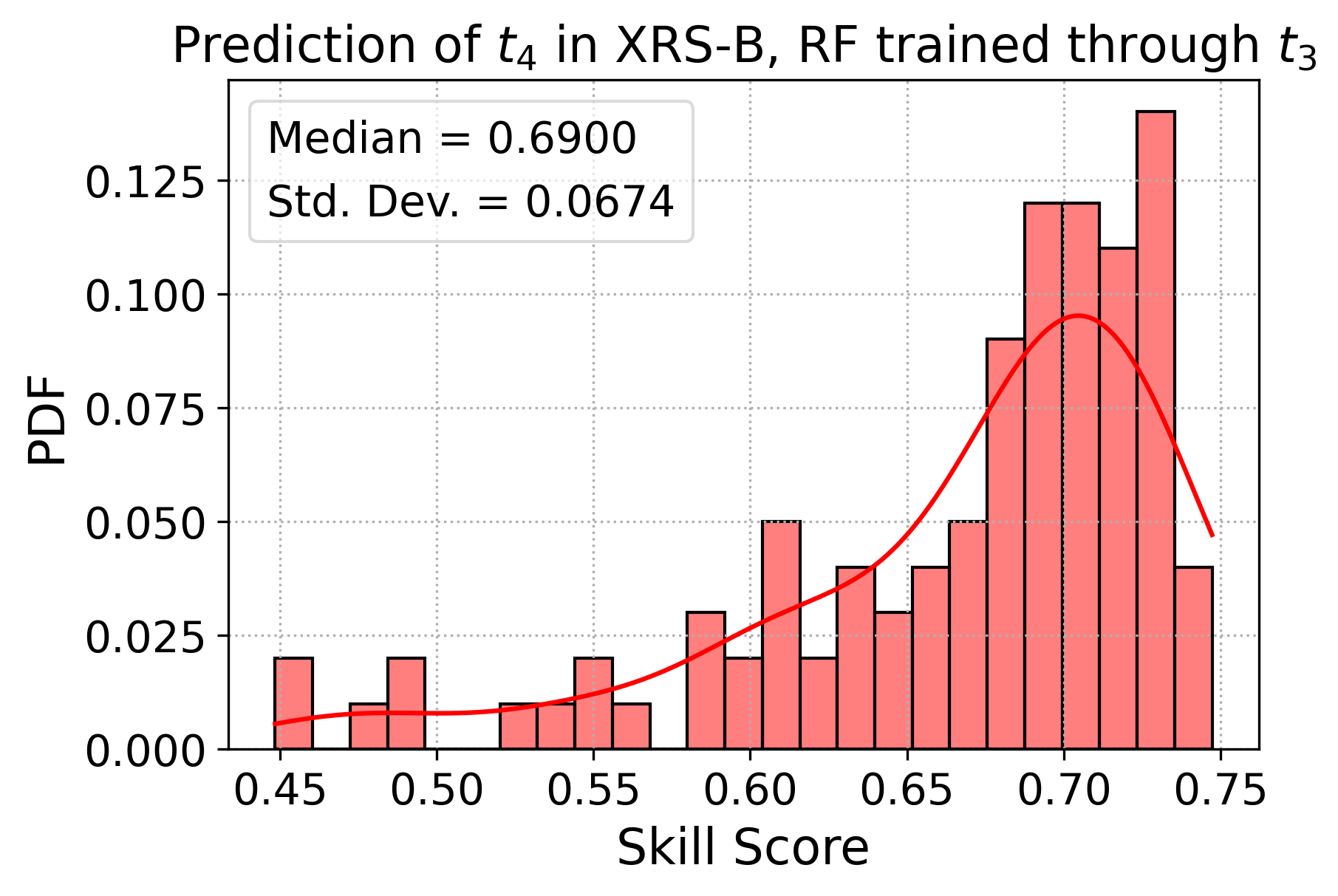}
\includegraphics[width=0.5\textwidth]{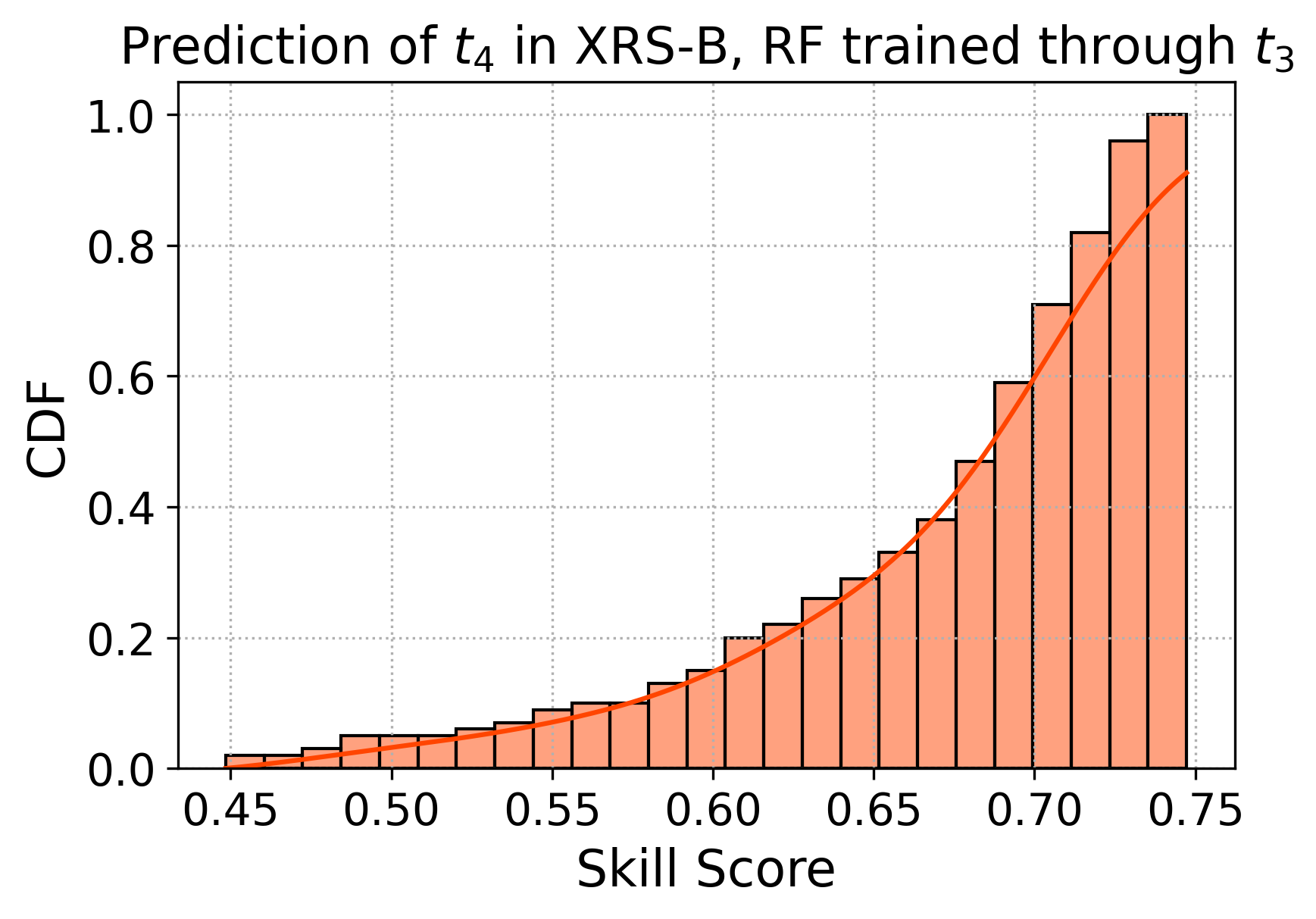}
\includegraphics[width=0.5\textwidth]{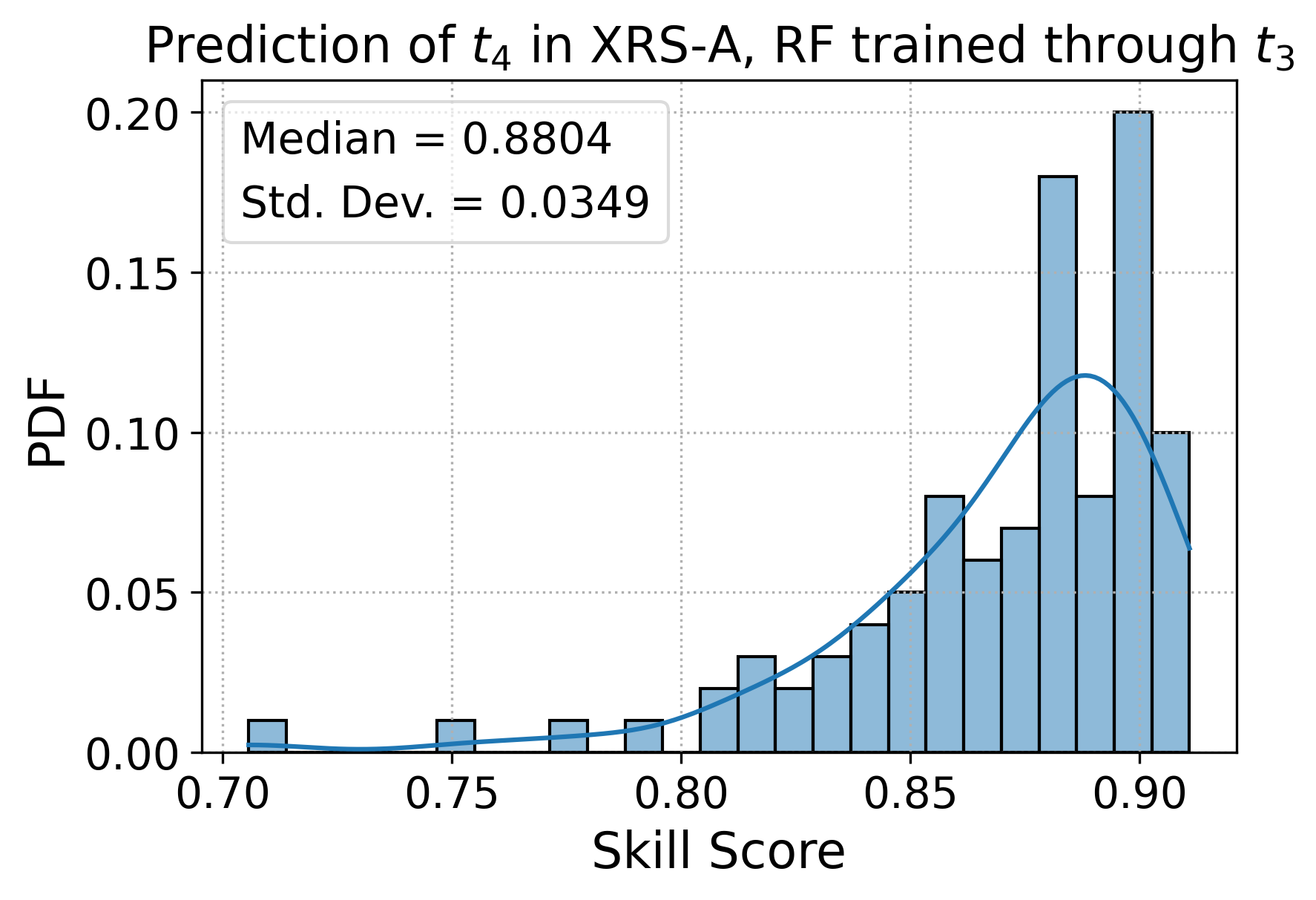}
\includegraphics[width=0.5\textwidth]{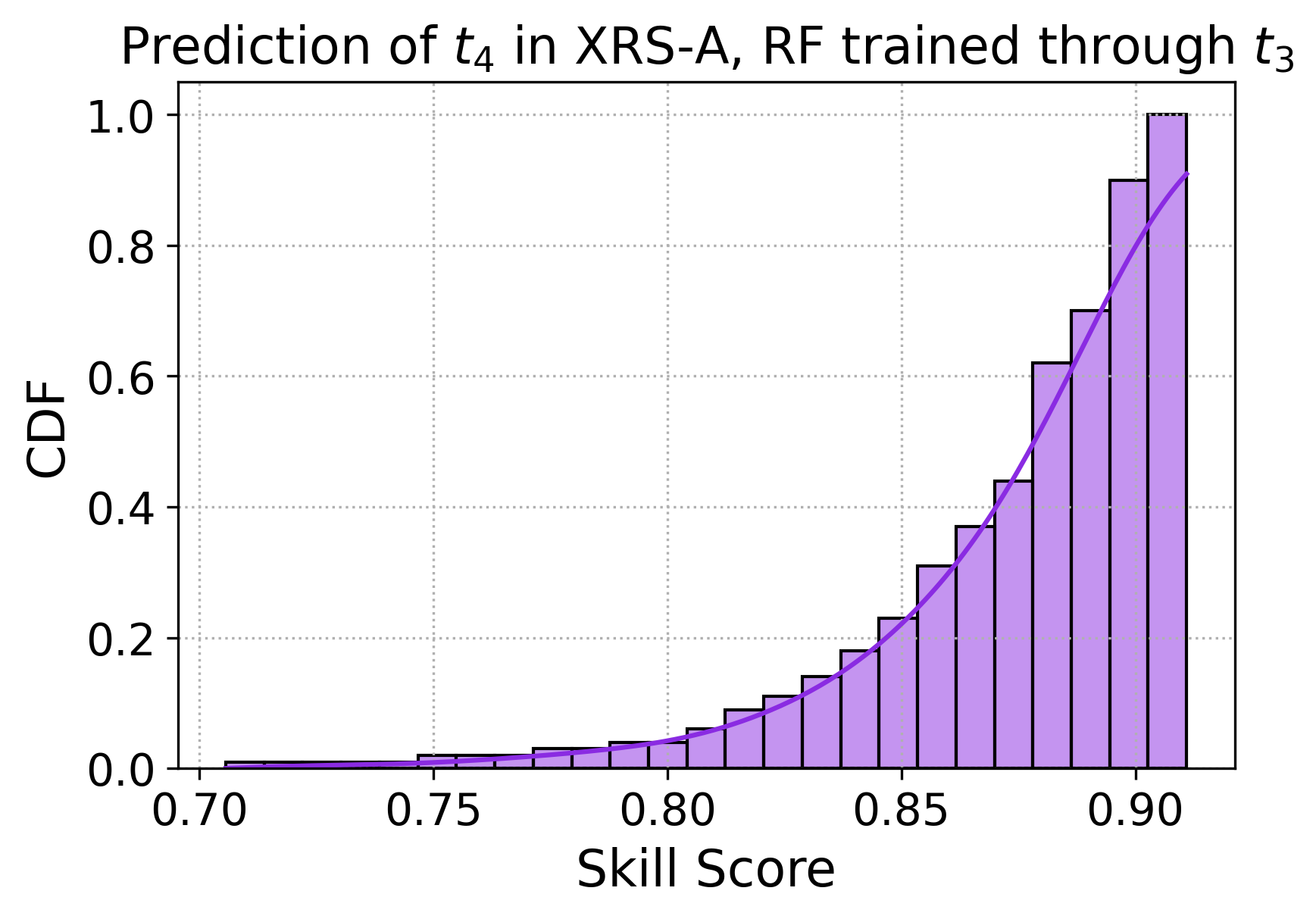}
\caption{Probability density functions and cumulative density functions, showing the distribution of the skill scores from a Monte Carlo simulation with 100 different random forests.  The top two rows show the results for the XRS-B and A channels, respectively, trained through time $t_{2}$.  The bottom two rows show the results trained through time $t_{3}$.  Trained through time $t_{2}$, the RF weakly, but not always, outperforms linear regression, and through time $t_{3}$, strongly outperforms linear regression.  \label{fig:pdfcdf}}
\end{figure*}

We should note, however, that the specific random forest in Section \ref{sec:data} was only one example, with one single seed, which may or may not be representative of a random forest in general.  To check the model, we therefore run a simple Monte Carlo simulation where we repeat these calculations for 100 different random forests.  We randomize the train-test split as well as the decision trees in the forests, and recalculate the skill scores.  The PDFs and cumulative distribution functions (CDFs) are shown in Figure \ref{fig:pdfcdf}.  In XRS-B, the random forest model provides a comparable prediction to the linear model when trained through time $t_{2}$ (median close to 0), and noticeably better than the linear model when trained through time $t_{3}$.  In XRS-A, it performs marginally better than the linear model at $t_{2}$ and markedly better at $t_{3}$.  The median values are slightly different from the first random forest example, but we find that the conclusion was roughly the same.  The random forest model provides a reasonable, though imperfect, forecast that statistically tends to slightly underestimate the values of $t_{4}$, with median error of less than 2 minutes.

\section{Conclusions}
\label{sec:conclusions}

This model is simple, lightweight, and accurate.  It can be run in Python to make direct predictions of the remaining duration, $t_{4} - t_{\mathrm{now}}$ at the current time $t_{\mathrm{now}}$, in an ongoing solar flare.  We have trained the data set using only parameters that are easily measured from GOES/XRS light curves, and which can be easily calculated in real-time.  The model generally performs well, with a skill score that outperforms a simpler linear regression model.  The error in the timing prediction is, on average, less than 2 minutes in both XRS channels.  The most important features for this forecasting are the timings of the start and peak as well as the fluence before and at the flare's peak.

We have additionally tested this model by using the features through time $t_{1}$ to predict those at $t_{2}$, $t_{3}$, and $t_{4}$ in each XRS channel.  As with the results presented in Section \ref{sec:data}, the model generally underestimates the values of the later times.  The method provides a reasonable prediction of $t_{2}$, but $t_{3}$ and $t_{4}$ are increasingly poorly constrained.  Plots of this can be found as a supplement in the code posted to Github for the interested reader.

We should note that, in principle, quasi-periodic pulsations (QPPs; \citealt{nakariakov2009}) could be measured from GOES/XRS light curves and that the period of QPPs are correlated with flare duration \citep{hayes2020}.  As noted by \citet{hayes2020}, however, the period changes during the course of a longer flare.  Furthermore, the signal-to-noise ratio needs to be relatively large to measure this period, which would exclude forecasting for smaller flares.  Finally, it is not clear that all flares exhibit QPPs: a sample of X-class flares detected QPPs in only about 80\% of the events \citep{simoes2015}.  In future work, it would be worthwhile to build QPP measurements into the model, particularly with newer GOES satellites which have better cadence.

While this model has been built with simplicity in mind, the addition of other data sets would likely improve the prediction.  The two XRS channels are generally sensitive only to plasma exceeding approximately 10 MK, and their ratio can be used to estimate temperature \citep{garcia1994}.  However, they do not actually measure the distribution of plasma at various temperatures, and the combination of this data with data from other instruments might strongly improve forecasts.  For example, the Extreme Ultraviolet Variability Experiment (EVE; \citealt{woods2012}) onboard the Solar Dynamics Observatory (SDO; \citealt{pesnell2012}) provides irradiance measurements of a wide range of spectral lines that form at different temperatures and heights in the solar atmosphere, which could potentially prove useful for these forecasts.  

Furthermore, while the random forest is lightweight and useful for these predictions, it is limited in predicting full time series.  Other machine learning algorithms, such as a recurrent neural network, could perhaps fare better in forecasting full time series for events.  The trade-off is that these would require training on full time series, rather than discrete values, and are significantly more computationally intensive.  A forecasting model of this type would be a major improvement to our current model, however.

\acknowledgments

J.W.R. was supported by a NASA Living With a Star Grant.  W.T.B. was supported by NASA's Hinode project through the National Research Council program.  The authors thank the referees and editor for their extremely thorough and helpful comments that have significantly improved this work.  The authors thank Douglas Drob, Dennis Socker, and Dave Siskind for helpful discussions in the development of this work, and what parameters to examine.  We thank Laura Hayes for help with SunPy usage, in particular GOES/XRS data in SunPy.  The authors also thank the SunPy collaboration for their efforts at making accessible tools for the solar community.

This work has made use of the following packages: \texttt{numpy} \citep{numpy}, \texttt{matplotlib} \citep{matplotlib}, \texttt{scipy} \citep{scipy}, \texttt{pandas} \citep{pandas}, \texttt{scikit-learn} \citep{scikit-learn}, \texttt{seaborn} \citep{seaborn}, and \texttt{sunpy} \citep{sunpy,sunpy_zenodo}.

\noindent\textbf{Data Availability Statement}. \\
The data and routines used to produce the results in this paper are available at \url{https://github.com/USNavalResearchLaboratory/flare_duration_forecasting}.  GOES/XRS data are provided by NOAA and were accessed through the SunPy package.

\bibliography{bib.bib}

\end{document}